\documentclass[showpacs,twocolumn,superscriptaddress]{revtex4-2}

\usepackage{graphicx}
\usepackage{physics}
\usepackage{wrapfig}
\usepackage{changepage}
\usepackage{amsmath,amssymb,amsthm,mathrsfs,amsfonts,dsfont,mathtools}
\usepackage[caption=false]{subfig}
\usepackage{epsfig}
\usepackage{color,xcolor}
\usepackage{adjustbox}
\usepackage{tabularx}
\usepackage{array}
\usepackage{multirow}
\usepackage{tabu}
\usepackage{bm}
\usepackage{enumerate}
\usepackage{hyperref}
\usepackage{newtxtext,newtxmath}
\usepackage{makecell} 
\usepackage{url}
\usepackage{booktabs}
\usepackage{nicefrac} 
\usepackage{makeidx}  
\usepackage{leftindex}
\usepackage{float}
\usepackage{algorithm}
\usepackage{algpseudocode}
\usepackage{cleveref}
\usepackage{mwe}

\usepackage{comment}
\usepackage{xurl}
\usepackage{subcaption}
\usepackage{booktabs}  
\usepackage{multirow}  
\usepackage{makecell}

\begin{document}
\title{Multi-channel convolutional neural quantum embedding}

\author{Yujin Kim}
\email{k.yujin2228@yonsei.ac.kr}
\affiliation{Department of Statistics and Data Science, Yonsei University, Seoul 03722, Republic of Korea}
\author{Changjae Im}
\affiliation{Department of Statistics and Data Science, Yonsei University, Seoul 03722, Republic of Korea}
\author{Taehyun Kim}
\affiliation{Department of Statistics and Data Science, Yonsei University, Seoul 03722, Republic of Korea}
\author{Tak Hur}
\affiliation{Department of Statistics and Data Science, Yonsei University, Seoul 03722, Republic of Korea}
\author{Daniel K. Park}
\email{dkd.park@yonsei.ac.kr}
\affiliation{Department of Statistics and Data Science, Yonsei University, Seoul 03722, Republic of Korea}
\affiliation{Department of Applied Statistics, Yonsei University, Seoul 03722, Republic of Korea}

\begin{abstract}
Classification using variational quantum circuits is a promising frontier in quantum machine learning. 
Quantum supervised learning (QSL) applied to classical data using variational quantum circuits involves embedding the data into a quantum Hilbert space and optimizing the circuit parameters to train the measurement process. In this context, the efficacy of QSL is inherently influenced by the selection of quantum embedding.
In this study, we introduce a classical-quantum hybrid approach for optimizing quantum embedding beyond the limitations of the standard circuit model of quantum computation (i.e., completely positive and trace-preserving maps) for general multi-channel data. 
We benchmark the performance of various models in our framework using the CIFAR-10 and Tiny ImageNet datasets and provide theoretical analyses that guide model design and optimization.
\end{abstract}

\maketitle

\section{Introduction}
\label{sec:intro}
Quantum machine learning (QML) is a promising approach to data processing and analysis beyond the capabilities of its classical counterparts~\cite{10.1038/s43588-022-00311-3, PhysRevLett.113.130503_QSVM, PhysRevLett.127.060503, doi:10.1126/science.abn7293}. 
Among several approaches, variational quantum algorithms (VQA)~\cite{cerezo2020variational, cong_quantum_2019, farhi2014quantumapproximateoptimizationalgorithm, Peruzzo2014, Bravo_Prieto_2023, verdon2019quantumhamiltonianbasedmodelsvariational, kim2025expressivity, Nguyen_2024, PERALGARCIA2024100619, e25020287, senokosov2024quantum} have become a practical choice for noisy intermediate-scale quantum (NISQ) devices~\cite{Preskill2018quantumcomputingin} due to their shallow circuit depth and adaptability to current quantum hardware. In particular, quantum classifiers based on VQAs are capable of solving supervised learning tasks
~\cite{suzuki_analysis_2020, VQASVM, blank2020quantum, park2021robust,PARK2020126422, grant_hierarchical_2018, PhysRevA.101.032308}.

A typical strategy in quantum supervised learning (QSL) via VQA is encoding classical data into quantum states as inputs to parameterized quantum circuits
and adjusting the parameters to optimize task performance.
However, QSL’s effectiveness is limited by the choice of quantum embedding. For instance, when training a quantum classifier to distinguish between two classes, the performance is bounded by the distinguishability between the resulting quantum states, which is commonly quantified by the trace distance~\cite{Helstrom1969, Bae_2015}. Importantly, completely positive trace-preserving (CPTP) maps cannot increase the trace distance between embedded quantum states~\cite{Nielsen:2011:QCQ:1972505}, placing a fundamental constraint on the performance after data embedding.

To overcome this limitation, neural quantum embedding (NQE), a novel approach, leverages classical neural networks to transform classical data and improve the distinguishability of quantum states~\cite{hur2023neural}.
This method maximizes the trace distance between embedded classes, thereby improving the classification performance in QSL. 
However, a systematic understanding of how architectural choices affect NQE’s performance remains lacking. These choices include the interface between the classical neural network and quantum circuit, training loss, and embedding circuit.

To address this problem, we propose a convolutional neural quantum embedding (CNQE) framework designed for general multi-channel data. 
Within this framework, we introduce three distinct interface structures between classical neural networks and quantum circuits, and analyze how different configurations influence the expressivity and separability of quantum states.
We further explore two training loss functions and investigate various embedding architectures.
Our study includes both theoretical formulations and extensive empirical evaluations of the proposed strategies.

Furthermore, we evaluate commonly used image classification datasets, such as CIFAR-10 and Tiny ImageNet, to demonstrate that CNQE can achieve high classification performance with reduced parameter complexity. This result highlights CNQE’s potential for practical, resource-efficient QML for real-world tasks.

The remainder of this paper is organized as follows.
Section~\ref{sec: Multi-channel CNQE} reviews the NQE framework.
Section~\ref{sec: Configurations}  introduces the proposed interface designs, loss functions, and embedding strategies, along with theoretical analyses.
Section~\ref{sec: Numerical Experiments} presents empirical evaluations of CNQE across various configurations and multi-channel datasets, analyzing the trace distance and classification accuracy to evaluate the impact of each configuration on the performance.
We also compare performance of CNQE with that of purely classical neural networks.
Finally, Section~\ref{sec: Conclusions and Discussions} concludes the paper with a discussion on future directions.

\section{Multi-channel CNQE}\label{sec: Multi-channel CNQE}

\begin{figure*}[t]
\centering
\includegraphics[width=0.8\textwidth]{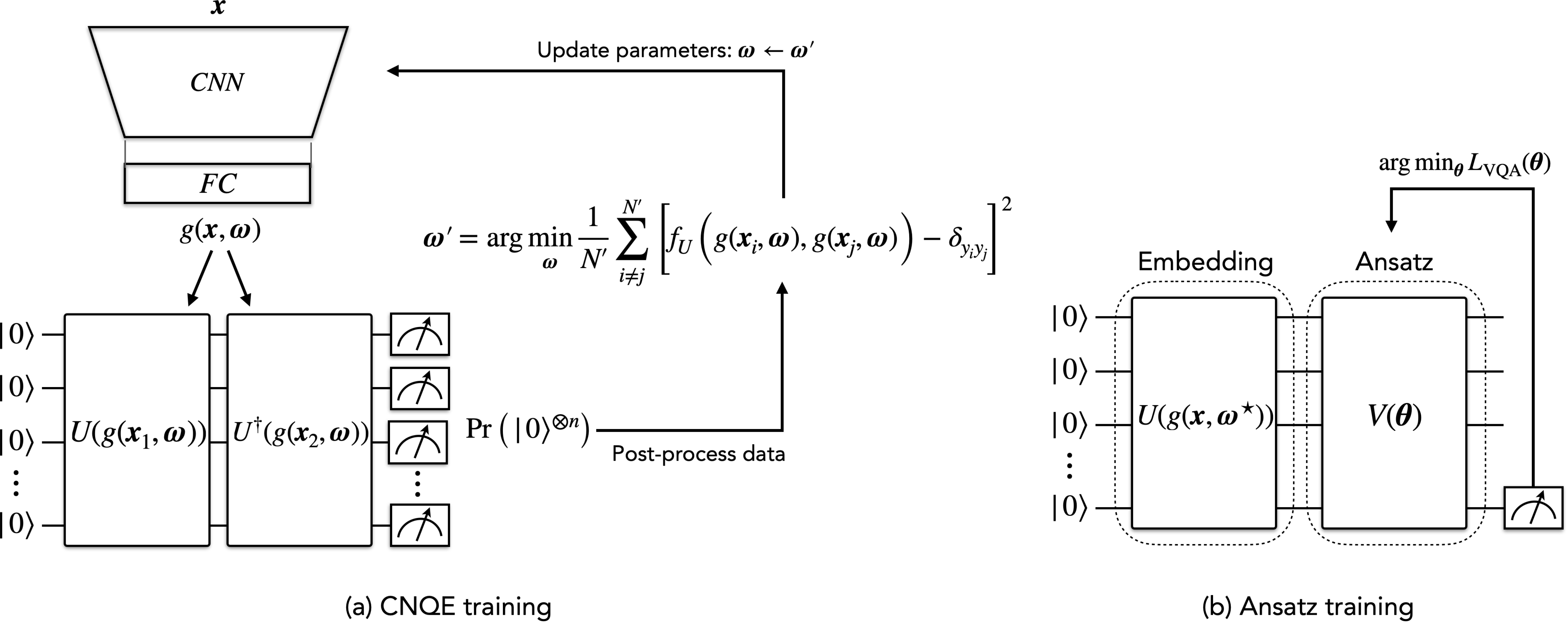}
\caption{\label{fig:Figure1} Overview of VQA through CNQE for multi-channel image data. (a) CNQE training: Finding optimal classical parameters $\boldsymbol{w}^\star$ to maximize the trace distance between the mean ensemble states with a unitary $U$ using similarity-based loss function $f_U$, by minimizing $L_{\mathrm{NQE}}$. (b) Ansatz training: After embedding the data, optimizing the POVM by adjusting quantum parameters $\theta$ to minimize $L_{\mathrm{VQA}}$.}
\end{figure*}

This section briefly reviews NQE and introduces its essential components for multi-channel image data.

NQE~\cite{hur2023neural} is a classical-quantum hybrid embedding scheme that learns the optimal data embedded states based on the theory of quantum state discrimination. 
Although originally proposed to improve the performance of binary QSL models implemented using VQAs,
it can also be applied to quantum kernel methods.

Let us recall a QSL for classical data. 
Given a dataset of $N$ labeled examples $\mathcal{D}=\{(\boldsymbol{x}_1,y_1),\cdots,(\boldsymbol{x}_N,y_N)\}$, 
where $\boldsymbol{x}\in\mathcal{X}=\mathbb{R}^{d}$ and
$y\in\mathcal{Y}=\mathbb{R}$ are drawn from an unknown distribution 
$\{\mathcal{X},\mathcal{Y}\}\subset\mathcal{P}$. 
The goal is to find a quantum map $h:\mathcal{X}\rightarrow\mathcal{Y}$ that produces a good estimation $y=h(\boldsymbol{x'})$ on the unseen data $\boldsymbol{x'}$ close to a true value $y'$ that is also drawn from $\mathcal{P}$.
To handle such classical datasets, QML methods embed classical data into quantum states using unitary map 
$\mathcal{E}: \mathcal{X}\rightarrow \mathcal{R},\, \mathcal{E}(\boldsymbol{x})=U(\boldsymbol{x})\rho_0U^\dagger(\boldsymbol{x})$, where $\mathcal{R}\in\mathcal{H}_q$ is a space of quantum state of a $q$-qubit Hilbert space, $U$ denotes a unitary embedding circuit, and $\rho_0$ is an initial quantum state.

Subsequently, for VQA, a parameterized unitary $V(\boldsymbol{\theta})$ with trainable parameters $\boldsymbol{\theta}$ define a hypothesis class
$\mathcal{H} = \left\{ h_{\boldsymbol{\theta}}(\boldsymbol{x}) = \mathrm{tr}\left[ V(\boldsymbol{\theta}) \mathcal{E}(\boldsymbol{x}) V^\dagger(\boldsymbol{\theta}) O \right] \right\},$
where each hypothesis $h_{\boldsymbol{\theta}}$ maps input $\boldsymbol{x}$ to a real-valued output by measuring the expectation of an observable O.
Hypothesis $h$ is optimized with respect to $\boldsymbol{\theta}$ by minimizing the empirical risk, which is defined as
\begin{equation}\label{eq: VQA loss}
    L_{\mathrm{VQA}}(\boldsymbol{\theta})
= \frac{1}{N} \sum_{i=1}^{N}
\,l(h_{\boldsymbol{\theta}}(\boldsymbol{x}_i),y_i)\,
\end{equation} 
where $l$ denotes binary classification loss.

By contrast, for the quantum kernel method, the kernel function $
k(\boldsymbol{x}, \boldsymbol{x}') = \mathrm{tr}\left[\mathcal{E}(\boldsymbol{x}) \mathcal{E}(\boldsymbol{x}')\right]
$ defines a decision function using the hypothesis class
$\mathcal{H} = \left\{ h(\boldsymbol{x}) = \sum_{i=1}^N \alpha_i k(\boldsymbol{x}, \boldsymbol{x}_i) \right\}.$
The coefficients $\{\alpha_i\}$ are obtained by minimizing empirical risk 
$(1/N)\sum_{i=1}^N l\left(h(\boldsymbol{x}_i), y_i\right) $.

For the binary classification, 
the error probability of distinguishing the embedded quantum states on the training dataset $\mathcal{D}$ is given by 
\begin{equation}\label{eq: VQA loss bound}
p_{\text{error}}(\mathcal{D})
\geq 
\frac{1}{2}-
D_{\mathrm{tr}}\left(q_+ 
\bar{\mathcal{E}}(\boldsymbol{x}_{+}), 
q_- \bar{\mathcal{E}}(\boldsymbol{x}_{-})\right)
\end{equation} 
This inequality is known as the Helstrom bound and characterizes the theoretical limit of binary quantum state discrimination~\cite{Helstrom1969, Bae_2015}.
Here, $\boldsymbol{x}_{\pm}\in\mathcal{X}_{\pm}$ are samples with labels $\mathcal{Y}=\pm 1$ and priors probabilities $q_+,q_-$.
The term $\bar{\mathcal{E}}(\boldsymbol{x}_{\pm})=\mathbb{E}_{\boldsymbol{x}_{\pm}\sim\mathcal{X}_{\pm}}\mathcal{E}(\boldsymbol{x}_{\pm})$ denotes the average embedded quantum state for each class.
$D_{\mathrm{tr}}(\cdot,\cdot)$ is the trace distance that quantifies the distinguishability between quantum states. This is defined as $D_{\mathrm{tr}}(\rho,\sigma)=\frac{1}{2}\|\rho-\sigma\|_1$, where the trace norm is $\|A\|_1=\mathrm{tr}\sqrt{A^{\dagger}A}$.

In the VQA setting, when a linear loss function is used, 
the VQA loss is lower-bounded by
\begin{equation}\label{eq: }
L_{\mathrm{VQA}}(\boldsymbol{\theta})
\geq 
\frac{1}{2}-
D_{\mathrm{tr}}\left(q_+ 
\bar{\mathcal{E}}(\boldsymbol{x}_{+}), 
q_- \bar{\mathcal{E}}(\boldsymbol{x}_{-})\right).
\end{equation} 
The MSE loss is tightly bounded by the linear loss via norm inequalities~\cite{hur2023neural}. These theoretical guarantees justify the use of the MSE loss, which has been widely adopted in machine learning. Accordingly, we use MSE as the loss function.

However, a quantum channel cannot increase the trace distance between two embedded states, because
\begin{equation*}
    D_{\mathrm{tr}}\left(q_+ 
\bar{\mathcal{E}}(\boldsymbol{x}_{+}), 
q_- \bar{\mathcal{E}}(\boldsymbol{x}_{-})\right)
\geq
D_{\mathrm{tr}}\left(\Phi(q_+ \bar{\mathcal{E}}(x_+)),\, \Phi(q_- \bar{\mathcal{E}}(x_-))\right)
\end{equation*}
where \( \Phi \) denotes an arbitrary CPTP map.
The NQE approach optimizes the input data embedding using a classical neural network with trainable parameters $\boldsymbol{w}$, producing modified inputs as 
\begin{equation*}
\boldsymbol{x}'_{\pm}=g(\boldsymbol{x}_{\pm},\boldsymbol{w})\,,
\end{equation*}
where $g(\boldsymbol{x}, \boldsymbol{w}) \in \mathbb{R}^p$ denotes the output vector of the neural network $g$.
$p$ corresponds to the number of features embedded in the quantum states, and its choice is arbitrary. In the following sections, we examine how both the architecture of the neural network and choice of $p$ affect classification performance.
This transformation increases the trace distance between class-averaged states, thereby lowering the bound of the VQA loss $L_{\mathrm{VQA}}$ and improving the binary classification performance.

$\boldsymbol{w}$ are trained by minimizing the following empirical risk over $N'$ sampled input pairs $\{(\boldsymbol{x}_i,y_i),(\boldsymbol{x}_j,y_j)\}\sim\mathcal{D} $:
\begin{equation}
\label{eq: NQE model}
L_{\mathrm{NQE}}(\boldsymbol{w})=
\frac{1}{N'}\sum_{i\neq j}^{N'}
\left[f_U\left(g(\boldsymbol{x}_i,\boldsymbol{w}),
g(\boldsymbol{x}_j,\boldsymbol{w})\right)-\delta_{y_iy_j}\right]^2.
\end{equation}
Here, $f_U$ denotes a CNQE loss function implemented by an embedding circuit $U$ that quantifies the similarity between the two quantum states. 
In this study, we consider the quantum state fidelity and Hilbert-Schmidt inner product to construct $f_U$. Further details of the loss functions are provided in~\ref{subsec: Configurations: Quantum Model}.

For classical network $g$, we employ a convolutional neural network (CNN) followed by fully connected (FC) layers. Accordingly, we name our method convolutional neural quantum embedding (CNQE).
An overview of the VQA framework with CNQE is shown in Fig.~\ref{fig:Figure1}. Figure~\ref{fig:Figure1}$(a)$ illustrates the training process of CNQE using the sampled input pairs. Figure~\ref{fig:Figure1}$(b)$ depicts the subsequent Ansatz training starting from the optimally embedded quantum states with the pre-trained CNQE parameters $\boldsymbol{w}$. 

\section{Configurations}\label{sec: Configurations}
\subsection{Classical-to-Quantum Interface}
\label{subsec: Configurations: Classical Model}
\begin{figure*}[t]
\centering
\includegraphics[width=\textwidth]{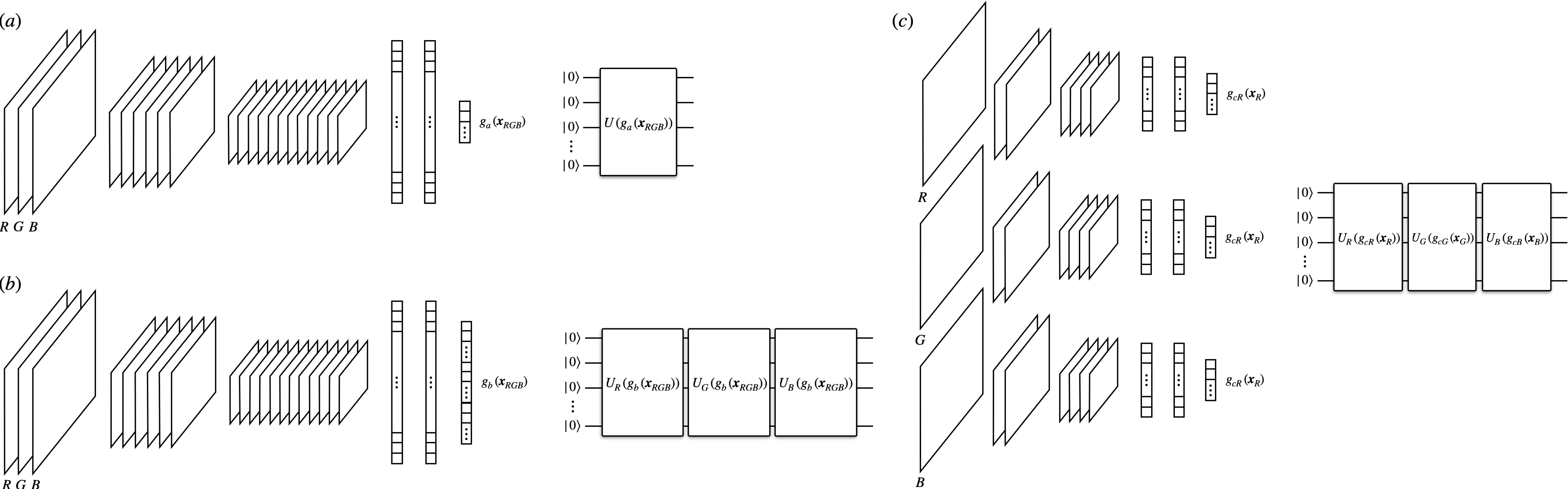}
\caption{\label{fig:CNN structures} Three classical neural network models for multi-channel convolutional NQE (CNQE). $(a)$ represents $g_a$, $(b)$ represents $g_b$, and $(c)$ represents $g_c$.}
\end{figure*}

The classical neural network $g$ in CNQE is an interface between the raw input data and quantum-embedded states.
The structure of $g$ should extract features that represent the input data effectively and enhance the distinguishability of the resulting quantum states in terms of trace distance. Simultaneously, computational overhead should be minimized. Therefore, the design of $g$ plays a critical role in balancing classification performance and efficiency.

Moreover, multi-channel image data involve interchannel dependencies, requiring additional design considerations compared to single-channel inputs. 
Considering these aspects, we introduce three different interface models $g$ and analyze their efficiency and expressivity. The proposed architecture is illustrated in Fig.~\ref{fig:CNN structures}. 

As shown in Fig.~\ref{fig:CNN structures}, we adopt CNNs with FC layers as the baseline for $g$, given their widespread use and effectiveness in processing multi-channel image data~\cite{Yamashita2018}. 
Model $g_a$ adopts a simple and naive architecture composed of a multi-channel CNN, followed by an FC hidden layer.
This model outputs a $p_{a}-$dimensional feature vector, where $p_a$ is determined by the input size required for the embedding circuit.
Input size scales with number of qubits in circuit.
Model $g_b$ retains the same CNN structure as in $g_a$ but expands the dimensionality of the output vector to $3p_{a}$, which is three times larger than that of $g_a$.
In model $g_c$, each channel passes through the same CNN structure and an FC hidden layer individually, using separate convolutional filters.
The outputs are then concatenated into a $3p_a-$dimensional vector.

Let $\boldsymbol{x}$ be a multi-channel input that can be expressed as a concatenation of individual channel inputs, $\boldsymbol{x} = [\boldsymbol{x}_1, \boldsymbol{x}_2, \ldots, \boldsymbol{x}_L]$. Without loss of generality, we consider RGB data, where $\boldsymbol{x} = [\boldsymbol{x}_R, \boldsymbol{x}_G, \boldsymbol{x}_B]$.
Mathematically, the output feature vector of each model is defined as 
\begin{align*}
\begin{split}
    &\bullet g_{a}(\boldsymbol{x}_{RGB})\\
    &\bullet g_{b}(\boldsymbol{x}_{RGB})\\
    &\bullet g_{c}(\boldsymbol{x}_{RGB})
    =
    [g_{cR}(\boldsymbol{x}_{R})
    ,\, g_{cG}(\boldsymbol{x}_{G})
    ,\, g_{cB}(\boldsymbol{x}_{B})]    
\end{split}
\end{align*}
Here, $g_i$ denotes each model and $g_i(\boldsymbol{x})$ represents its corresponding output feature vector.
In the case of model $g_c$, the output is obtained by concatenating the outputs of the three separate models $g_{cR}, g_{cG}$, and $g_{cB}$. 
Each model processes single input channels $\boldsymbol{x}_R$, $\boldsymbol{x}_G$, and $\boldsymbol{x}_B$.
Although we illustrate the method using RGB channels, the design of models $g_a$, $g_b$, and $g_c$ can be generalized to any number of channels. In general, the output dimensionality of $g_b$ and $g_c$ scale with the number of input channels, resulting in an output of size (number of channels) $\times p_a$.

The embedding circuit for each model is designed based on the dimensionality of the output vector.
For a unit embedding circuit that requires $x$ parameters, model $g_a$ employs a single embedding layer to generate a $p_a-$dimensional output.
By contrast, models $g_b$ and $g_c$ produce a $3p_a-$dimensional output vector and utilize three embedding circuit layers.
The details of the embedding circuit architecture are discussed in Section ~\ref{subsec: Configurations: Embedding Circuit}.

In models $g_a$ and $g_b$, the interchannel dependencies are captured through the CNN structure. Moreover, the resulting features exhibit nonlinearity owing to the use of the activation functions. Model $g_b$, with a higher dimensional output, enables richer feature extraction, and sequentially embeds these features into the deeper layers of the quantum circuit.
By contrast, model $g_c$ treats each channel independently and linearly concatenates its outputs. Thus, its classical part lacks the nonlinearity associated with interchannel relationships.
However, nonlinearity can still be introduced during the quantum embedding process because the embedding circuit is nonlinear with respect to the input. This allows the model to capture complex relationships, even in the absence of explicit classical modeling of interchannel dependencies. This aspect is analyzed in detail in this section.

To assess and compare the efficiency and expressive capacity of the proposed models, we first analyze the number of trainable parameters involved in the CNQE process and then
examine the expressivity of the quantum circuits after embedding.

The number of trainable parameters in the interface model $g$ is a key metric for evaluating the model efficiency.
Table~\ref{tab: CNN parameters} lists the number of trainable parameters used in our experiments, assuming that each quantum layer has a $p_a-$dimensional output vector. In the table, $n_c$ denotes the number of input channels and $n_f$ denotes the output dimension of the CNN block, which also corresponds to the flattened input dimension of the first FC layer. The parameter count for CNN  is computed under fixed conditions, including the kernel size and stride. 
Further architectural details and parameter counts for each classical model are provided in Appendix~\ref{Append: Details of the Classical Model Used in CNQE Training}.
\begin{table}[h]
\setlength{\tabcolsep}{7pt}
\renewcommand{\arraystretch}{1.3}
\centering
\begin{tabular}{|c|c|c|}
\hline
&CNN&FC\\ \hline
$g_a$& $90{n_c}^2+6n_c$& ${n_f}^2+{n_f}+{n_f}p_a+p_a$\\ \hline
$g_b$& $90{n_c}^2+6n_c$& ${n_f}^2+{n_f}+n_c{n_f}p_a+n_c p_a$
\\ \hline
$g_c$& $96n_c$& ${n_f}^2/n_c+{n_f}+{n_f}p_a+n_c p_a$
\\ \hline
\end{tabular}
\caption{Number of trainable parameters in each interface model, separated into CNN and FC components.
The convolution kernel size is $3\times3$, the pooling kernel size is $4\times4$, and the stride is set to 4.}
\label{tab: CNN parameters}
\end{table}

As shown in Table~\ref{tab: CNN parameters}, models $g_a$ and $g_b$ share the same CNN architecture, resulting in an identical number of CNN parameters, whereas $g_c$ employs a smaller CNN with fewer parameters, that is, $\mathrm{CNN}_a = \mathrm{CNN}_b > \mathrm{CNN}_c$.
For FC layers, $g_b$ has the largest number of parameters because of its multi-channel structure, followed by $g_a$ and $g_c$, leading to the relation $\mathrm{FC}_b > \mathrm{FC}_a > \mathrm{FC}_c$.
Here, $\mathrm{CNN}_{\,i}$ and $\mathrm{FC}_{\,i}$ denote the number of parameters in the CNN and FC layers of model $g_i$, respectively.
Based on these observations, the total number of trainable parameters across the models follows the order 
\begin{equation}\label{eq: classical param order}
g_b > g_a > g_c.
\end{equation}

Although the parameter count provides a measure of model efficiency, it does not directly reflect how expressive the resulting quantum state is.
To evaluate the expressivity of each model, we analyze the quantum states produced after embedding.
Expressivity refers to the complexity of the function class that a parametric model can represent. In QML, expressivity is often characterized by a Fourier analysis, which captures the intrinsic properties of quantum states and operations~\cite{schuld2021effect, kim2025expressivity}. 
When the input data $g(\boldsymbol{x})$ are embedded using a quantum circuit $U$,
the resulting quantum state is given by
\begin{equation*}
    U(g(\boldsymbol{x}))
    |\boldsymbol{0}_n\rangle,
\end{equation*}
where we assume that the initial state is $|\boldsymbol{0}_n\rangle = |0\rangle^{\otimes n}$  for an $n$-qubit circuit.

\[
[U(g_a(\boldsymbol{x}_{RGB}))|0_n\rangle]_j = \sum_{\omega_a \in \Omega_a} (c_{\omega_a})_j \exp(i \omega_a(\boldsymbol{x}_{RGB})),
\]

For each model, the $k$-th component of the resulting quantum state can be expressed in terms of its Fourier sum as
\begin{align}\label{eq: fourier analysis}
    \begin{split}
        [U(g_a(\boldsymbol{x}_{RGB}))|\boldsymbol{0}_n\rangle]_k
        &=\sum_{\omega_a\in\Omega_a}(c_{\omega_a})_k
        \exp(i\omega_a(\boldsymbol{x}_{RGB}))\,,\\
        [U(g_b(\boldsymbol{x}_{RGB}))|\boldsymbol{0}_n\rangle]_k
        &=\sum_{\omega_b\in\Omega_b}(c_{\omega_b})_k
        \exp(i\omega_b(\boldsymbol{x}_{RGB}))\,,\\
        [U(g_c(\boldsymbol{x}_{RGB}))|\boldsymbol{0}_n\rangle]_k
        &=\sum_{\omega_c\in\Omega_c}(c_{\omega_c})_k
        \exp(i\omega_c(\boldsymbol{x}_{RGB}))\,,
    \end{split}
\end{align}
where $\omega_j(\boldsymbol{x}_{RGB})$ for $j=\{a,b,c\}$ denotes the frequency component in the set $\Omega_j$ and $(c_{\omega_j})_k$ represents the corresponding coefficient for the $k$-th amplitude component.
Each frequency component \( \omega_j \in \Omega_j \) is a linear combination of elements of the embedded input vector \( g_j(\boldsymbol{x}_{RGB}) \). 
More precisely,
\begin{align}\label{eq: fourier analysis frequency}
    \begin{split}
        &\quad\omega_a(\boldsymbol{x}_{RGB})
        \in \left\{\boldsymbol{h}_{a}\cdot g_a(\boldsymbol{x}_{RGB})\right\}
        ,\\
        &\quad\omega_b(\boldsymbol{x}_{RGB})
        \in \left\{\boldsymbol{h}_{b}\cdot g_b(\boldsymbol{x}_{RGB})\right\}
        ,\\        
        &\quad\omega_c(\boldsymbol{x}_{RGB})
        \in \left\{\boldsymbol{h}_{c}\cdot g_c(\boldsymbol{x}_{RGB})\right\}
        \\
        &=  \left\{\boldsymbol{h}_{cR}\cdot g_{cR}(\boldsymbol{x}_{R})
        + \boldsymbol{h}_{cG}\cdot g_{cG}(\boldsymbol{x}_{G})
        + \boldsymbol{h}_{cB}\cdot g_{cB}(\boldsymbol{x}_{B})\right\}
        ,\\
    \end{split}
\end{align}
where each \( \boldsymbol{h}_j \in \mathbb{R}^p \) denotes a possible coefficient vector in a linear combination that defines the frequency component \( \omega_j \).
The set of vectors is determined by the manner in which the components of the input vector $g_j(\boldsymbol{x}_{RGB})$ are assigned to specific locations within a unitary circuit.
The Fourier coefficient $(c_{\omega_j})_k$ is determined by the nonparametric structure of the quantum circuit, such as the arrangement of CNOT gates and choice of Pauli rotation gates.
The details of the Fourier analysis in Eqs.~\eqref{eq: fourier analysis} and \eqref{eq: fourier analysis frequency} as well as the explicit form of $(c_{\omega_j})_k$ are provided in Appendix~\ref{Append: Fourier Analysis of Embedded Quantum State}.

This analysis reveals that each component of the quantum state can be expressed as the sum of exponential functions, where the frequencies are linear combinations of the input features.
This implies that the input is embedded in a quantum state through a nonlinear transformation defined by the quantum circuit.
Although model $g_c$ processes each input channel independently at the classical level, 
interchannel information can still be captured during the quantum embedding step. 
This is enabled by the nonlinear nature of the quantum circuit with a repeated encoding structure.
Thus, all of our classical–quantum interface models ultimately incorporate interchannel information through quantum embedding.
These observations together show that model $g_c$ uses the fewest trainable parameters in a classical network, minimizing computational cost.
This indicates that model $g_c$ can remain computationally efficient while being sufficiently expressive to capture complex patterns in multi-channel data.

\subsection{Loss Functions}
\label{subsec: Configurations: Quantum Model}
We now explore the possible formulations of the CNQE loss function $f_U$. As defined in Eq.\eqref{eq: NQE model}, CNQE employs a similarity-based loss. Therefore, quantum kernel functions\cite{blank2020quantum, PARK2020126422} are natural candidates for constructing $f_U$. In this study, we consider two similarity measures: quantum state fidelity and the Hilbert-Schmidt inner product.

The fidelity-based loss is defined as follows: 
\begin{equation}\label{eq: quantum model 1}
    {f_U}_{\mathrm{fid}}(\boldsymbol{x}_1,\boldsymbol{x}_2)
    =|\langle \boldsymbol{0}_n|U^\dagger(\boldsymbol{x}_1)U(\boldsymbol{x}_2)|\boldsymbol{0}_n\rangle|^2.
\end{equation}
This assumes that the quantum circuit is initialized in state $|\boldsymbol{0}_n\rangle$.
This quantity measures the overlap between the output states $U(\boldsymbol{x}_1)|\boldsymbol{0}_n\rangle$ and $U(\boldsymbol{x}_2)|\boldsymbol{0}_n\rangle$, effectively quantifying the similarity between the first-column vectors of the unitaries. Hence, designing $U$ such that its first column is easily tunable enhances the model's expressivity for CNQE training.

The Hilbert-Schmidt-based loss is given by
\begin{equation}
\label{eq: quantum model 2}
    {f_U}_{\mathrm{HS}}(\boldsymbol{x}_1,\boldsymbol{x}_2)=
    \frac{1}{2^n}\mathrm{tr} \left(U^\dagger(\boldsymbol{x}_1)U(\boldsymbol{x}_2)\right),
\end{equation}
which quantifies the similarity between the unitary operators $U(\boldsymbol{x}_1)$ and $U(\boldsymbol{x}_2)$ via their Hilbert-Schmidt inner product.
This metric captures the global similarity across a full orthonormal basis and does not depend on a fixed input state.
Thus, it is well suited for settings in which the initial state is arbitrary or mixed.
Ensemble-based quantum platforms, such as nuclear magnetic resonance (NMR) and electron spin resonance (ESR), which operate with mixed-state inputs, benefit from this formulation.
Notably, deterministic quantum computation with a one-qubit (DQC1) model, which is widely used in ensemble quantum computing~\cite{doi:10.1073/pnas.94.5.1634,doi:10.1098/rsta.2011.0352,NMRSimulator,Kyungdeock2015,Lu2016}, naturally outputs the normalized trace of a unitary operator. Therefore, the Hilbert-Schmidt loss aligns well with such architectures and enhances the applicability of the CNQE in practical QML tasks~\cite{kim2025expressivity,liu2025neuralquantumembeddingdeterministic}.

\subsection{Embedding Circuit}
\label{subsec: Configurations: Embedding Circuit}
\begin{figure}[t]
\centering
\includegraphics[width=\columnwidth]{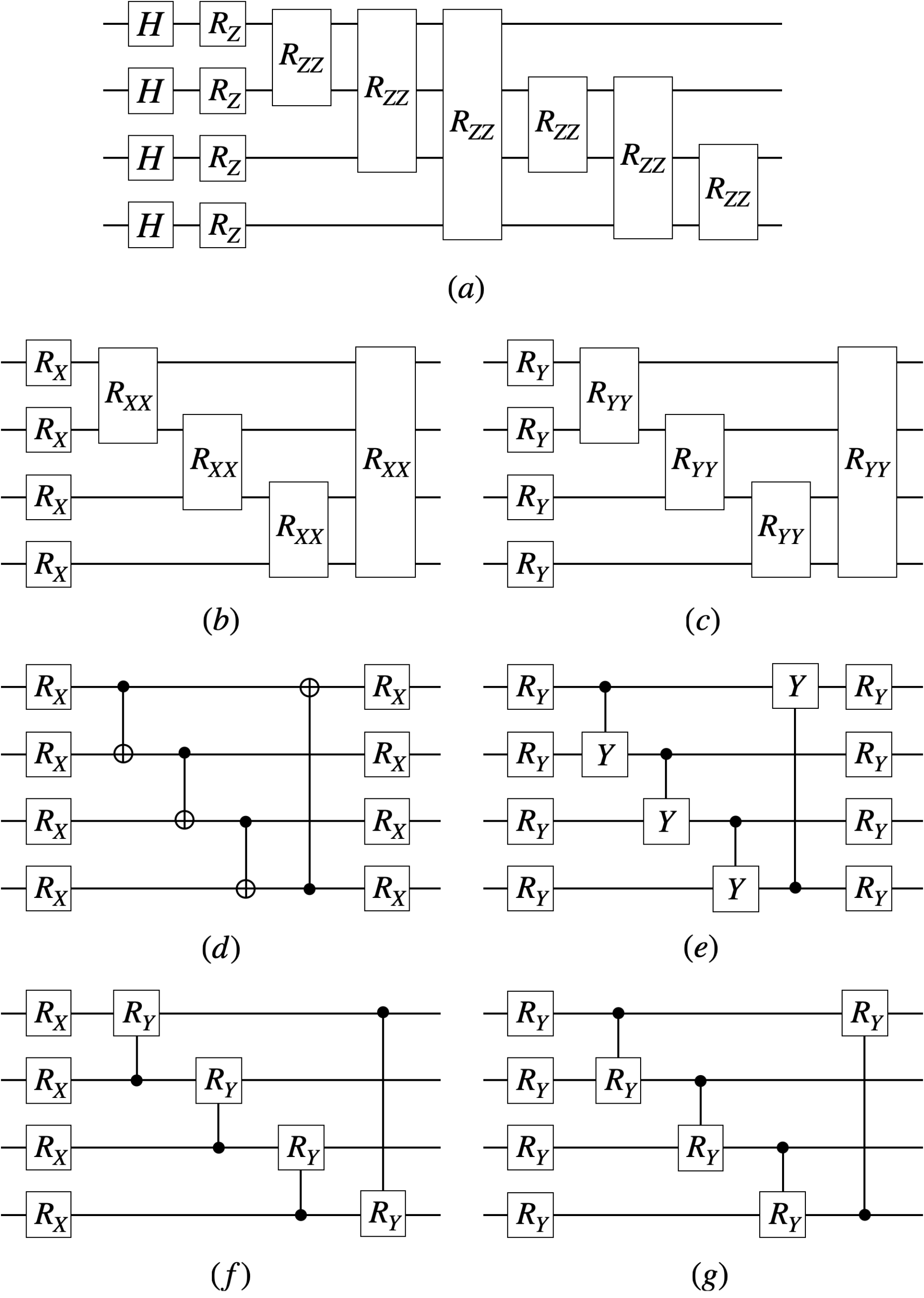}
\caption{\label{fig: embedding circuit} Embedding circuits considered in this study. Gates $R_X,R_Y,R_Z,R_{XX},R_{YY}$, and $R_{ZZ}$ are parameterized.}
\end{figure}

The design of the embedding circuit $U$ determines how the quantum states are represented in Hilbert space. 
This influences the trainability and performance of NQE~\cite{holmes_2022,Sim_expressibility}. This also affects the ability to learn a POVM that effectively distinguishes between embedded states~\cite{Lee_2024}.

In this study, we explore seven types of embedding circuit units, as shown in Fig.~\ref{fig: embedding circuit}.
Each unit corresponds to a single-circuit layer designed for data embedding in the proposed model architecture.
These circuits can be categorized into four types, based on their structural similarities.
$g_a$ embeds the output vector using a single circuit layer, whereas $g_b$ and $g_c$ sequentially embed an output vector three times larger across the three circuit layers. For these models, the circuit is constructed by stacking three units belonging to the same structural category.

The first circuit $U_1(\boldsymbol{\theta})$ shown in Fig.~\ref{fig: embedding circuit}(a) shows the $ZZ$-feature embedding~\cite{Havlicek2019}. It is expressed as 
\begin{equation}
U_1(\boldsymbol{\theta})
=\exp(i\sum_i\phi_{\{i\}}(\boldsymbol{\theta})Z_i+i\sum_{i<j}\phi_{\{i,j\}}(\boldsymbol{\theta})Z_iZ_j)H^{\otimes n}\,,
\end{equation}
where $\phi(\boldsymbol{\theta})$ is the data-encoding function with input $\boldsymbol{\theta}$ and $H$ represents the Hadamard gate. 
In CNQE, $\boldsymbol{\theta}$ corresponds to the data $g(\boldsymbol{x})$ processed by the classical neural network $g$.
For the encoding map, we use $\phi_{\{i\}}(\boldsymbol{\theta})=\boldsymbol{\theta}_i$ and $\phi_{\{i,j\}}(\boldsymbol{\theta})=\boldsymbol{\theta}_{k}\,(k\in\{n+1,\cdots,n(n-1)/2\})$. 
Accordingly, the unit circuit requires ${n(n+1)}/2$ parameters for an $n$-qubit system.
The $ZZ$ feature map is a specific implementation of the instantaneous quantum polynomial time (IQP) feature map~\cite{PhysRevLett.117.080501} known for its classically hard-to-simulate quantum computations.  
The $ZZ$ feature map encodes data nonlinearly via coefficients $\phi$, employing diagonal gates in the Pauli $Z$-basis, while considering all combinations of qubit entanglements.
The embedding scheme of models $g_b$ and $g_c$ involves sequentially stacking $U_1(\boldsymbol{\theta})$ circuits, each receiving a distinct component of the output vector as the input.
The resulting operation is defined as follows:
$U_{\mathrm{ZZ}}(\boldsymbol{\theta}^{(1)},\boldsymbol{\theta}^{(2)},\boldsymbol{\theta}^{(3)})
=U_1(\boldsymbol{\theta}^{(3)}),
U_1(\boldsymbol{\theta}^{(2)})
U_1(\boldsymbol{\theta}^{(1)})$.

In the second category, corresponding to Fig.~\ref{fig: embedding circuit}(b) and Fig.~\ref{fig: embedding circuit}(c),
we consider a modified version of the $ZZ$ feature embedding. 
This approach uses two-qubit gates that involve rotations with respect to the Pauli $X$ and $Y$ operators.
It employs a circular entanglement scheme in which each qubit interacts with its nearest neighbors, including the connection between the first and last qubits.
The embedding unitaries are expressed as
\begin{align}
\begin{split}
U_2(\boldsymbol{\theta})
&=\exp(i\sum_i\phi_{\{i\}}(\boldsymbol{\theta})
X_i+i\sum_{i}\phi_{\{i,i'\}}(\boldsymbol{\theta})X_iX_{i'})\,,    \\
U_3(\boldsymbol{\theta})
&=\exp(i\sum_i\phi_{\{i\}}(\boldsymbol{\theta})
Y_i+i\sum_{i}\phi_{\{i,i'\}}
(\boldsymbol{\theta})Y_iY_{i'})\,,    
\end{split}
\end{align}
where $\phi_{\{i\}}(\boldsymbol{\theta})=\boldsymbol{\theta}_i$ and $\phi_{\{i,i'\}}(\boldsymbol{\theta})=\boldsymbol{\theta}_{i+n}$, satisfying $i'=i+1\,(\mathrm{mod\,n})$.
Each unit takes $2n$ parameters as inputs for an $n$-qubit system.
The three-layer embedding circuit is constructed as $U_2(\boldsymbol{\theta})$ and $U_3(\boldsymbol{\theta})$ as
\begin{equation}
    U_{\mathrm{NC}}(\boldsymbol{\theta}^{(1)},\boldsymbol{\theta}^{(2)},\boldsymbol{\theta}^{(3)})
=U_2(\boldsymbol{\theta}^{(3)})
U_3(\boldsymbol{\theta}^{(2)})
U_2(\boldsymbol{\theta}^{(1)})\,,
\end{equation}
where each unit encodes the corresponding input.
Each circuit shares the same structure but operates on a different Pauli basis; therefore, the data are encoded in noncommutative bases across the layers.
We refer to this configuration as an NC (noncommutative) feature map.

In the third group, the embedding circuits shown in Fig.~\ref{fig: embedding circuit}(d) and Fig.~\ref{fig: embedding circuit}(e) are adapted from Circuit 10 in Ref.~\cite{Sim_expressibility} with modifications.
Circuit 10 follows a hardware-efficient architecture~\cite{kandala_hardware-efficient_2017}, which incorporates a circular entanglement structure.
The circuits are defined as
\begin{align}
\begin{split}
U_4(\boldsymbol{\theta})
&=\exp(i\sum_i\phi'_{\{i\}}(\boldsymbol{\theta})X_i)
\prod_{i}\mathrm{CX}_{ii'}
    \exp(i\sum_i\phi_{\{i\}}(\boldsymbol{\theta})X_i)\,,\\
U_5(\boldsymbol{\theta})
&=\exp(i\sum_i\phi'_{\{i\}}(\boldsymbol{\theta})Y_i)
\prod_{i}\mathrm{CY}_{ii'}
    \exp(i\sum_i\phi_{\{i\}}(\boldsymbol{\theta})Y_i)\,,
\end{split}
\end{align}
where $\phi_{\{i\}}(\boldsymbol{\theta})=\boldsymbol{\theta}_i$, $\phi'_{\{i\}}(\boldsymbol{\theta})=\boldsymbol{\theta}_{i+n}$, and each unit requires $2n$ parameters.
In the two-qubit gates $\mathrm{CX}_{i,i'}$ and $\mathrm{CY}_{i,i'}$, the subscripts indicate that the $i$th qubit is the control and $i'$th qubit is the target, with $i' = i+1\, (\mathrm{mod\,n})$.
In the three-layer embedding setting, the circuit is constructed as
\begin{equation}
    U_{\mathrm{NC}10}(\boldsymbol{\theta}^{(1)},\boldsymbol{\theta}^{(2)},\boldsymbol{\theta}^{(3)})
=U_4(\boldsymbol{\theta}^{(3)})
U_5(\boldsymbol{\theta}^{(2)})
U_4(\boldsymbol{\theta}^{(1)})\,,
\end{equation}
which is referred to as the NC10 feature map.

The fourth group consists of the circuits shown in Fig.\ref{fig: embedding circuit}(f) and Fig.\ref{fig: embedding circuit}(g), which are motivated by the rotational components of the Lorentz transformation around the $x$ and $y$ axes in 4D spacetime.
This concept is briefly explored in~\cite{toffano:hal-03042373}.
The matrix forms of the Lorentz rotations are expressed as
\begin{align*}
    \begin{split}
    J_x&=\begin{pmatrix}
1 & 0 & 0 & 0 \\
0 & 1 & 0 & 0 \\
0 & 0 & \cos\theta & -\sin\theta \\
0 & 0 & \sin\theta & \cos\theta
\end{pmatrix},\quad
    J_y=\begin{pmatrix}
1 & 0 & 0 & 0 \\
0 & \cos\theta & 0 & \sin\theta \\
0 & 0 & 1 & 0 \\
0 & -\sin\theta & 0 & \cos\theta
\end{pmatrix},\\
    J_z&=\begin{pmatrix}
1 & 0 & 0 & 0 \\
0 & \cos\theta & -\sin\theta & 0 \\
0 & \sin\theta & \cos\theta & 0 \\
0 & 0 & 0 & 1
\end{pmatrix},   
\end{split}
\end{align*}
where $J_x$, $J_y$, and $J_z$ represent rotations around the $x$, $y$, and $z$ axes, respectively.
These rotations can be implemented using the two-qubit gates $CRY_{ji}$, $CRY_{ij}$, ${R_{XX+YY}}_{ij}$ defined as
\begin{align}
    \begin{split}
    CRY_{ji}(2\theta)&=\begin{pmatrix}
1 & 0 & 0 & 0 \\
0 & 1 & 0 & 0 \\
0 & 0 & \cos\theta & -\sin\theta \\
0 & 0 & \sin\theta & \cos\theta
\end{pmatrix},\\
    CRY_{ij}(2\theta)&=\begin{pmatrix}
1 & 0 & 0 & 0 \\
0 & \cos\theta & 0 & \sin\theta \\
0 & 0 & 1 & 0 \\
0 & -\sin\theta & 0 & \cos\theta
\end{pmatrix},\\
    {R_{XX+YY}}_{ij}(2\theta,\pi/2)&=\begin{pmatrix}
1 & 0 & 0 & 0 \\
0 & \cos\theta & -\sin\theta & 0 \\
0 & \sin\theta & \cos\theta & 0 \\
0 & 0 & 0 & 1
\end{pmatrix}.   
\end{split}
\end{align}
Using this framework, we construct circuits composed of single-qubit rotation layers around the $x$- and $y$-axes
followed by the corresponding Lorentz rotation layer around the same axis, which is defined as
\begin{align}
\begin{split}    
U_6(\boldsymbol{\theta})&=\prod_{i} CRY_{i'i} (2\phi'_{\{i\}}(\boldsymbol{\theta}))
\exp(i\sum_i\phi_{\{i\}}(\boldsymbol{\theta})X_i)\,,\\
U_7(\boldsymbol{\theta})&=\prod_{i} CRY_{ii'} (2\phi'_{\{i\}}(\boldsymbol{\theta}))
\exp(i\sum_i\phi_{\{i\}}(\boldsymbol{\theta})Y_i)\,,
\end{split}
\end{align}
where $\phi_{\{i\}}(\boldsymbol{\theta})=\boldsymbol{\theta}_i, \phi'_{\{i\}}(\boldsymbol{\theta})=\boldsymbol{\theta}_{i+n}$ and  $i' = i+1\, (\mathrm{mod\,n})$.
Each unit has $2n$ input parameters.
The three-layer embedding circuit is then given by
\begin{equation}
U_{\mathrm{NCL}}(\boldsymbol{\theta}^{(1)},\boldsymbol{\theta}^{(2)},\boldsymbol{\theta}^{(3)})
=U_6(\boldsymbol{\theta}^{(3)})
U_7(\boldsymbol{\theta}^{(2)})
U_6(\boldsymbol{\theta}^{(1)})\,.
\end{equation}
This structure is referred to as a NCL (noncommutative Lorentz) feature map.

\section{Numerical Experiments}\label{sec: Numerical Experiments}
We evaluate CNQE under various configurations, different interface models, loss functions, and embedding circuits. Experiments are conducted on binary classification tasks using CIFAR-10~\cite{krizhevsky2009learning} and Tiny ImageNet~\cite{krizhevsky2009learning}, because of their size and class diversity (see also~\cite{senokosov2024quantum} for another study applying quantum models to CIFAR-10).
Three binary classification tasks are selected for evaluation: Airplane–Automobile and Frog–Ship from CIFAR-10, and School Bus–Maypole from Tiny ImageNet. Further details regarding the datasets and training procedures are provided in Appendix~\ref{Append: Experimental Setup: Data and Training Details}.

The main simulation results are summarized in Table~\ref{tab: combined_results} and were obtained on a 4-qubit system. To examine scalability, we additionally performed experiments on 6- and 8-qubit systems using the CIFAR-10 frog–ship pair and the Tiny ImageNet class pair. As reported in Appendix~\ref{Append: Extended Experiments with 8-Qubit Simulation}, these results show that performance remains stable as the number of qubits and trainable parameters increase, indicating that CNQE-QCNN maintains effectiveness without degradation from the larger state space.
Performance is assessed using two metrics. The trace distance is used to evaluate the embedding quality of CNQE, and the classification accuracy is measured after training a QCNN ansatz on the embedded states.
Supplementary results for precision, recall, and F1-score across all binary classification experiments are provided in Appendix~\ref{Append: Precision, Recall, and F1-Score for Binary Classification Experiments}. These additional evaluations are consistent with the accuracy-based conclusions, further reinforcing the robustness of our findings.

In this analysis, first, we examine the overall relationship between trace distance and classification accuracy. Next, we investigate how the trace distance varies depending on the interface model, quantum loss function, and embedding circuit. We also analyze the impact of different interface models on classification accuracy. 
In addition, we compare our models with purely classical neural networks. We further consider models that employ classical dimensionality reduction methods instead of CNQE. Finally, we evaluate the performance and noise robustness of our CNQE-based model relative to these baselines.

\newcolumntype{M}[1]{>{\centering\arraybackslash}m{#1}} 
\begin{table*}[!]
    \centering
    \begin{tabular}{M{3cm} | M{1.5cm} M{2.0cm} | M{2.0cm} M{2.0cm} | M{2.0cm} M{2.0cm}}    
        \toprule
        \multirow{2}{*}{\textbf{Data}} &
        \multirow{2}{*}{\makecell{\textbf{Interface} \\ \textbf{Model}}} & 
        \multirow{2}{*}{\makecell{\textbf{Embedding} \\ \textbf{Circuit}}} & 
        \multicolumn{2}{c|}{\textbf{Fidelity-based}} & 
        \multicolumn{2}{c}{\textbf{Hilbert-Schmidt-based}} \\
        \cmidrule{4-7}
        & & & \makecell{\textbf{Trace} \\ \textbf{Distance}} & \makecell{\textbf{Accuracy} (\%)} 
          & \makecell{\textbf{Trace} \\ \textbf{Distance}} & \makecell{\textbf{Accuracy} (\%)} \\
        \bottomrule
        \vspace{0.1cm}
        \multirow{15}{*}{\makecell{CIFAR-10\\(Airplane–Automobile)}} 
        &\multirow{7}{*}{$g_a$}
        & $(a)$ & 0.623 & 83.2 $\pm$ 0.25 & 0.583 & 83.0 $\pm$ 0.63 \\
        && $(b)$ & 0.595 & 83.2 $\pm$ 0.87 & \textbf{0.654} & \textbf{87.4} $\pm$ 0.20 \\
        && $(c)$ & 0.614 & \textbf{83.7} $\pm$ 0.25 & \textbf{0.679} & \textbf{85.9} $\pm$ 0.20 \\
        && $(d)$ & 0.606 & 81.0 $\pm$ 0.32 & \textbf{0.675} & 84.3 $\pm$ 0.25 \\
        && $(e)$ & 0.621 & 82.5 $\pm$ 0.00 & 0.615 & 83.1 $\pm$ 0.20 \\
        && $(f)$ & 0.599 & 81.9 $\pm$ 0.73 & 0.464 & 81.7 $\pm$ 0.40 \\
        && $(g)$ & \textbf{0.647} & 83.3 $\pm$ 0.25 & 0.425 & 84.0 $\pm$ 0.00 \\
        \cmidrule{2-7}
        &\multirow{4}{*}{$g_b$}
        & $U_{\mathrm{ZZ}}$ & \textbf{0.645} & \textbf{85.7} $\pm$ 0.60 & 0.574 & 79.2 $\pm$ 1.47 \\
        && $U_{\mathrm{NC}}$ & \textbf{0.649} & 84.0 $\pm$ 0.00 & 0.617 & 84.7 $\pm$ 0.25 \\
        && $U_{\mathrm{NC10}}$ & 0.622 & 83.7 $\pm$ 0.40 & 0.645 & 82.7 $\pm$ 1.63 \\
        && $U_{\mathrm{NCL}}$ & 0.596 & 77.7 $\pm$ 0.40 & 0.551 & 79.9 $\pm$ 0.20 \\
        \cmidrule{2-7}
        &\multirow{4}{*}{$g_c$}
        & $U_{\mathrm{ZZ}}$ & 0.602 & 83.3 $\pm$ 0.98 & 0.511 & \textbf{85.6} $\pm$ 1.32 \\
        && $U_{\mathrm{NC}}$ & 0.527 & 79.8 $\pm$ 2.54 & 0.560 & 79.6 $\pm$ 1.83 \\
        && $U_{\mathrm{NC10}}$ & 0.578 & \textbf{85.8} $\pm$ 0.40 & 0.535 & 77.5 $\pm$ 1.34 \\
        && $U_{\mathrm{NCL}}$ & 0.537 & 81.1 $\pm$ 1.46 & 0.514 & 82.8 $\pm$ 0.25 \\
        \bottomrule
        
        \vspace{0.1cm}
        \multirow{15}{*}{\makecell{CIFAR-10\\(Frog–Ship)}} 
        &\multirow{7}{*}{$g_a$}
        & $(a)$ & 0.832 & \textbf{95.7} $\pm$ 0.40 & 0.813 & \textbf{94.2} $\pm$ 0.25 \\
        && $(b)$ & 0.823 & 94.6 $\pm$ 0.49 & \textbf{0.858} & 92.9 $\pm$ 0.20 \\
        && $(c)$ & 0.815 & 93.0 $\pm$ 0.00 & 0.847 & 92.6 $\pm$ 3.07 \\
        && $(d)$ & 0.817 & \textbf{95.3} $\pm$ 2.93 & 0.854 & 91.9 $\pm$ 0.49 \\
        && $(e)$ & 0.792 & 94.0 $\pm$ 0.00 & 0.853 & 92.7 $\pm$ 0.25 \\
        && $(f)$ & 0.793 & 94.9 $\pm$ 0.20 & 0.851 & 92.0 $\pm$ 0.00 \\
        && $(g)$ & 0.847 & 94.0 $\pm$ 0.00 & 0.815 & 89.4 $\pm$ 0.20 \\
        \cmidrule{2-7}
        &\multirow{4}{*}{$g_b$}
        & $U_{\mathrm{ZZ}}$ & \textbf{0.851} & \textbf{95.3} $\pm$ 0.25 & \textbf{0.872} & 93.5 $\pm$ 1.79 \\
        && $U_{\mathrm{NC}}$ & 0.823 & 92.3 $\pm$ 0.40 & \textbf{0.878} & \textbf{96.5} $\pm$ 0.00 \\
        && $U_{\mathrm{NC10}}$ & \textbf{0.848} & 95.0 $\pm$ 0.00 & 0.857 & 91.2 $\pm$ 3.42 \\
        && $U_{\mathrm{NCL}}$ & \textbf{0.859} & 94.5 $\pm$ 0.00 & 0.786 & \textbf{96.2} $\pm$ 1.63 \\
        \cmidrule{2-7}
        &\multirow{4}{*}{$g_c$}
        & $U_{\mathrm{ZZ}}$ & 0.744 & 92.5 $\pm$ 0.00 & 0.728 & 91.9 $\pm$ 1.36 \\
        && $U_{\mathrm{NC}}$ & 0.751 & \textbf{95.3} $\pm$ 0.40 & 0.809 & 93.4 $\pm$ 0.49 \\
        && $U_{\mathrm{NC10}}$ & 0.792 & 94.3 $\pm$ 0.40 & 0.809 & 91.4 $\pm$ 0.20 \\
        && $U_{\mathrm{NCL}}$ & 0.755 & 94.2 $\pm$ 0.40 & 0.785 & 91.9 $\pm$ 1.16 \\
        \bottomrule
        
        \vspace{0.1cm}
        \multirow{15}{*}{\makecell{Tiny ImageNet\\(School Bus-Maypole)}} 
        &\multirow{7}{*}{$g_a$}
        & $(a)$ & \textbf{0.857} & \textbf{94.0} $\pm$ 0.00 & \textbf{0.845} & \textbf{92.8} $\pm$ 1.47 \\
        && $(b)$ & \textbf{0.851} & \textbf{95.7} $\pm$ 0.40 & \textbf{0.834} & \textbf{91.4} $\pm$ 0.49 \\
        && $(c)$ & 0.831 & 91.5 $\pm$ 0.00 & 0.779 & 90.6 $\pm$ 0.49 \\
        && $(d)$ & 0.829 & \textbf{94.0} $\pm$ 0.00 & 0.821 & 88.2 $\pm$ 1.17 \\
        && $(e)$ & 0.783 & 90.2 $\pm$ 0.40 & 0.820 & 91.2 $\pm$ 0.40 \\
        && $(f)$ & 0.832 & 93.4 $\pm$ 0.49 & 0.759 & 90.4 $\pm$ 0.49 \\
        && $(g)$ & 0.796 & 92.0 $\pm$ 0.00 & 0.821 & 88.6 $\pm$ 0.49 \\
        \cmidrule{2-7}
        &\multirow{4}{*}{$g_b$}
        & $U_{\mathrm{ZZ}}$ & 0.812 & 91.8 $\pm$ 0.98 & 0.827 & 90.8 $\pm$ 0.40 \\
        && $U_{\mathrm{NC}}$ & 0.812 & 89.0 $\pm$ 0.00 & \textbf{0.834} & 91.0 $\pm$ 0.00 \\
        && $U_{\mathrm{NC10}}$ & 0.801 & 91.6 $\pm$ 0.80 & \textbf{0.854} & \textbf{91.6} $\pm$ 0.49 \\
        && $U_{\mathrm{NCL}}$ & \textbf{0.845} & 92.8 $\pm$ 0.75 & 0.816 & 77.2 $\pm$ 0.40 \\
        \cmidrule{2-7}
        &\multirow{4}{*}{$g_c$}
        & $U_{\mathrm{ZZ}}$ & 0.643 & 87.8 $\pm$ 0.40 & 0.654 & 91.2 $\pm$ 0.40 \\
        && $U_{\mathrm{NC}}$ & 0.654 & 90.0 $\pm$ 0.00 & 0.704 & 91.0 $\pm$ 0.00 \\
        && $U_{\mathrm{NC10}}$ & 0.626 & 89.4 $\pm$ 1.20 & 0.674 & 87.0 $\pm$ 1.10 \\
        && $U_{\mathrm{NCL}}$ & 0.636 & 83.8 $\pm$ 0.75 & 0.633 & 86.0 $\pm$ 0.00 \\
        \bottomrule
    \end{tabular}
    \caption{
    Simulation results for CNQE and subsequent QCNN training across three data pairs. Each pair includes 15 models combining different interface models, embedding circuits, and loss functions. Median trace distance and mean classification accuracy (with standard deviation) are reported. Trace distance is computed from five independent CNQE runs, and accuracy is measured by training QCNN five times on the state with median trace distance. Top three results for each loss function are highlighted in bold.
    }
    \label{tab: combined_results}
\end{table*}

\subsection{Trace Distance–Accuracy Correlation}
\label{subsec: Numerical Experiments: Trace Distance–Accuracy Correlation}

\begin{figure}[h]
\centering
\includegraphics[width=0.95\columnwidth]{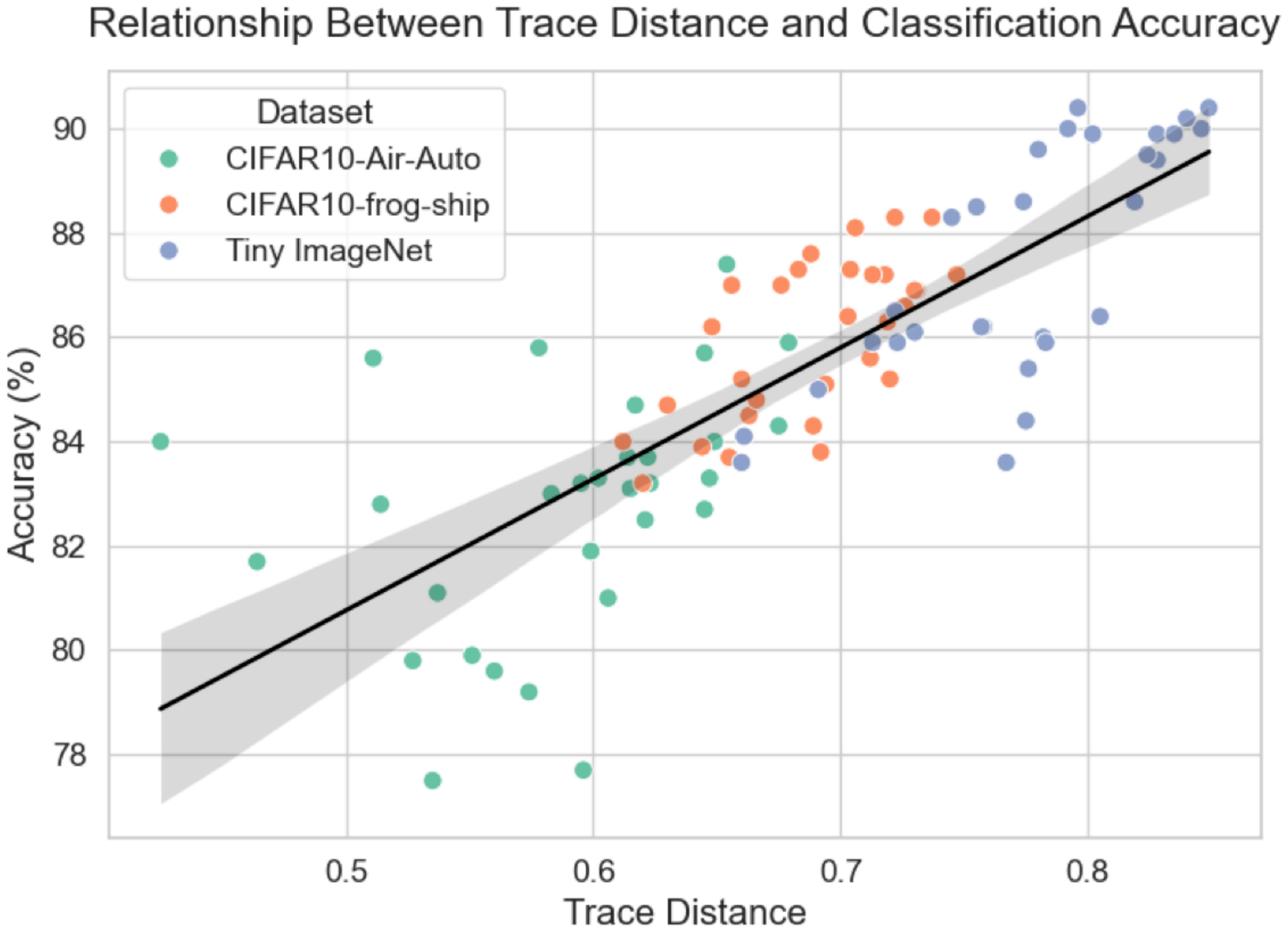}
\caption{Relationship between trace distance and classification accuracy across datasets. The solid black line represents the linear regression fit with 95\% confidence interval shaded.}
\label{fig: correlation - TDvsCA} 
\end{figure}

Figure~\ref{fig: correlation - TDvsCA} shows the correlation between trace distance and classification accuracy across all configurations. Each data point corresponds to a configuration, with the median trace distance and mean classification accuracy computed over five independent runs. Tables~\ref{tab: combined_results} presents the corresponding numerical values.

The figure demonstrates a strong positive correlation supported by Pearson’s $r = 0.7926\,(p < 0.001)$ and Spearman’s $\rho = 0.8184\,(p < 0.001)$. These results indicate tendency of larger trace distances to correspond to an improved classification performance. This reflects the core principles of the CNQE strategy. However, other factors such as optimization variability and dataset-specific characteristics can also influence the final performance and should be considered when interpreting the results.

\subsection{Effect of Model Architecture on Trace Distance}
\label{subsec: Numerical Experiments: Effect of Model Architecture on Trace Distance}

We present the individual effects of the interface model type, loss function, and embedding circuit on the resulting trace distances, as summarized in Appendix~\ref{Append: Trace Distance Statistics by Model Component} and Table~\ref{tab:trace_distance_summary}. 
We use 450 trace distance samples collected from 5 independent runs for each combination of 3 datasets, 2 loss functions, and 15 model configurations defined by the interface model types and embedding circuits.

To assess whether the differences between components are statistically significant, we conduct pairwise Welch’s t-tests on the final trace distance values for each quantity of interest because each group contains at least 30 samples.
Furthermore, to account for the increased risk of type I errors arising from multiple hypothesis testing, the $p$-values obtained from the pairwise comparisons are adjusted using the Bonferroni correction.

To analyze the interface model types, each group ($g_a$, $g_b$, and $g_c$) contains 210, 120, and 120 samples, respectively. The Bonferroni-corrected $p$-values indicate that the difference between $g_a$ and $g_b$ is not statistically significant ($p = 1.000$), whereas comparisons involving $g_c$ ($g_a$ vs.\ $g_c$ and $g_b$ vs.\ $g_c$) are statistically significant ($p < 0.001$ in both cases), with $g_c$ yielding smaller trace distances.

The effect of the loss function is evaluated based on 225 samples per group. The resulting $p$-value of 0.8444 indicates no statistically significant difference between the two loss functions.

The effect of the embedding circuit is analyzed separately for the $g_a$ group and the combined $g_b$ and $g_c$ groups. This separation is due to the differences in the embedding circuit types depending on the interface model. In both groups, all pairwise Welch’s t-tests yield Bonferroni-corrected $p$-values of $1.000$, indicating no statistically significant differences among the ansatz types. Each ansatz type is evaluated using 30 samples from the $g_a$ group and 60 samples from the $g_b$ and $g_c$ groups.

Overall, the results suggest that the loss function and embedding circuit have no statistically significant effect on the trace distance.  
In contrast, the interface model $g_c$ consistently produces significantly smaller trace distances than models $g_a$ and $g_b$, with no significant difference observed between $g_a$ and $g_b$.

\subsection{Effect of Interface Models on Classification Accuracy}
\label{subsec: Numerical Experiments: Effect of Interface Models on Classification Accuracy}

Section~\ref{subsec: Numerical Experiments: Effect of Model Architecture on Trace Distance} showed that only the interface model led to statistically significant differences in the trace distance among the three architectural components. Hence, we fix the loss function and embedding circuit and evaluate the classification accuracy across three different interface models, which is the primary objective of CNQE.

As discussed in Section~\ref{subsec: Numerical Experiments: Trace Distance–Accuracy Correlation}, the trace distance and classification accuracy have a strong positive correlation. However, the correlation is imperfect (i.e., the Pearson and Spearman coefficients are below 1), indicating that additional factors may affect the classification performance. Therefore, we conduct further simulations with a larger sample size to enable a more robust comparison of the classification accuracy across the interface models.

This analysis is conducted on the Frog-Ship pair from CIFAR-10 dataset, which yields the highest accuracy among the three datasets. All models are trained using the Hilbert–Schmidt loss. For the embedding circuit, we use $(b)$ for the interface model $g_a$, and $U_{\mathrm{NC}}$ for both $g_b$ and $g_c$. In each case, CNQE is trained to optimize quantum embedding, followed by QCNN training for classification. Each configuration is repeated 30 times with randomly initialized parameters. The training hyperparameters, including the optimizer type, learning rate, and number of iterations, are the same as those in Tab. ~\ref{tab: combined_results}.
Table~\ref{tab: results_cifar10_frog_ship_further_sim} summarizes these results.

\begin{table}[H]
    \centering
    \setlength{\tabcolsep}{5pt}
    \renewcommand{\arraystretch}{1.1}
    \begin{tabular}{c}
    \begin{minipage}[h]{0.5\textwidth}
        \centering
        \begin{tabular}{c|c|c|c}
            \toprule
            \textbf{\makecell{Interface \\ Model}} & \textbf{\makecell{Embedding \\ Circuit}} & \textbf{\makecell{Trace \\ Distance}} & \textbf{Accuracy (\%)} \\
            \midrule
            $g_a$ & $(b)$     & $0.827 \pm 0.054$ & $93.75 \pm 1.50$ \\\hline
            $g_b$ & $U_{\mathrm{NC}}$  & $0.840 \pm 0.049$ & $93.93 \pm 1.95$ \\\hline
            $g_c$ & $U_{\mathrm{NC}}$  & $0.773 \pm 0.073$ & $92.67 \pm 1.84$ \\
            \bottomrule
        \end{tabular}
        \caption*{\small (a)}
    \end{minipage}
    \vspace{0.5cm}
    \\
    \begin{minipage}[h]{0.48\textwidth}
        \centering
        \begin{tabular}{c|c|c}
            \toprule
            & \multicolumn{2}{c}{\textbf{$p$-value (Bonferroni)}} \\
            \cline{2-3}
            & \textbf{Trace Distance} & \textbf{Accuracy (\%)} \\
            \hline
            $g_a$ vs.\ $g_b$ & 1.0000 & 1.0000 \\\hline
            $g_a$ vs. $g_c$ & 0.0057 & 0.0463 \\\hline
            $g_b$ vs.\ $g_c$ & 0.0003 & 0.0366 \\
            \bottomrule
        \end{tabular}
        \caption*{\small (b) }
    \end{minipage}
    \end{tabular}

    \caption{
        Performance comparisons among interface models on the Frog–Ship pair from CIFAR-10. (a) Average trace distance and classification accuracy over 30 runs. (b) Bonferroni-corrected $p$-values for pairwise comparisons.
    }
    \label{tab: results_cifar10_frog_ship_further_sim}
\end{table}

We analyze the pairwise statistical differences among the interface models with respect to both trace distance and classification accuracy. The trace distance results are consistent with the findings in Section~\ref{subsec: Numerical Experiments: Trace Distance–Accuracy Correlation}. While $g_a$ and $g_b$ show no statistically significant differences, $g_c$ exhibits a significantly smaller trace distance.

A similar trend is observed in the classification accuracy, where $g_c$ performs comparably or slightly better than $g_a$ and $g_b$. However, the differences in the mean accuracy values are smaller, and the Bonferroni-corrected $p$-values are closer to the conventional significance threshold (0.05), indicating weaker statistical significance. This suggests that while $g_c$ shows a statistically significant smaller trace distance than the other models, this difference is less clear in the classification performance. In other words, the clear gap observed in the trace distance is less evident in terms of classification accuracy. 

This result suggests that $g_c$, despite having the fewest trainable parameters in the CNQE stage, has comparable classification accuracy. This supports the assessment that $g_c$is an efficient interface model.

\subsection{Performance Comparison with Classical Baseline Models}
\label{subsec: Performance Comparison with Classical Baseline Models}

To evaluate the effectiveness of our CNQE-based QCNN models, we compare their performance with that of purely classical neural networks. 
Based on the architecture used for 4-qubit simulations, we construct the corresponding classical baseline models. We retain the classical portion of the CNQE model and replace QCNN with a classical neural network.
To ensure a fair comparison, classical models are designed with the number of trainable parameters similar to or greater than that of the QCNN (15 parameters); specifically, the parameter counts range from 13 to 50, depending on the model architecture.

To account for the differences in output dimensionality across the interface models, with $g_b$ and $g_c$ producing three times more output than $g_a$, the subsequent classical layers are adjusted accordingly. All models are trained using the Adam optimizer with a learning rate of 0.01 for 50 epochs and a batch size of 25. A cross-entropy loss function is employed. 
Details of the classical baseline model architecture and their classification accuracies on the three datasets are provided in Appendix~\ref{Append: Summary of Classical Baseline Model Performance}, Table~\ref{tab: classical_model_comparison}.

We compare the performance of the quantum models in Table~\ref{tab: combined_results} with that of classical baseline models. Among the classical baselines, only the simplest architectures comprising a single linear layer (e.g., linear (8,2) or linear (24,2)) demonstrate a performance comparable to that of CNQE-based QCNNs. 
Certain architectures are overly simplistic, resulting in poor learning performance, with the accuracy remaining near 50\%.
Nevertheless, these classical models require more trainable parameters (18 or 50) than QCNN (15).

We focus on the $g_c$ interface model, which exhibits the highest parameter efficiency. The best classical baseline models outperform certain quantum models across CIFAR-10 (Automobile–Airplane), CIFAR-10 (Frog–Ship), and Tiny ImageNet, with mean accuracies of $81.3\pm2.1$, $91.9\pm1.7$, and $84.0\pm2.6$, respectively. However, the highest-performing quantum models achieves superior results on the same datasets, with accuracies of $85.8\pm0.4$, $95.3\pm0.4$, and $91.2\pm0.4$, respectively. Notably, the best classical models requires approximately three times more trainable parameters (50) than the QCNN (15).

Similar trends are observed for $g_a$ and $g_b$, where CNQE-based QCNNs consistently achieve comparable or superior performance to the best classical baselines despite having a similar or smaller number of parameters.

In conclusion, the interface model $g_c$ achieves competitive classification accuracy with the fewest trainable parameters, demonstrating the best parameter efficiency among all configurations.

\subsection{Performance Comparison with Autoencoder-Based Dimensionality Reduction}\label{subsec: Performance Comparison with Autoencoder-Based Dimensionality Reduction}

To further assess the effectiveness of CNQE, we compared its performance with that of an autoencoder-based dimensionality reduction approach under noiseless simulations. 

In CNQE, the embedding is optimized directly through a neural network, whereas in the autoencoder approach the input is first compressed into a lower-dimensional latent representation, which is then embedded and processed by the variational circuit for binary classification. While a comparison with PCA could also be considered, autoencoders are adopted here as they capture nonlinear and more general representations beyond the linear structures preserved by PCA. This comparison highlights the contribution of CNQE to representation learning in our setting.

For this comparison, the 4-qubit CNQE–QCNN model was used as the reference. Two autoencoder models were trained, each corresponding to $g_a$ and $g_c$ models. As the embedding ansatz for the QCNN models, $(b)$ and $U_{\mathrm{NC}}$ were employed, respectively. The results are summarized in Table~\ref{tab: autoencoder_comparison_noisy_simulation}, and the detailed architectures and experimental settings are provided in Appendix~\ref{Append: Performance Comparison with Autoencoder-Based Dimensionality Reduction}.

\newcolumntype{C}[1]{>{\centering\arraybackslash}p{#1}}

\begin{table*}[!ht]
\centering
\renewcommand{\arraystretch}{1.2}
\begin{tabular}{C{3.2cm}|C{2.5cm}|C{2.5cm}|C{2cm}|C{2.5cm}|C{2cm}}
\toprule
\textbf{Data} & \textbf{\makecell{Corresponding\\Interface Model}} & 
\multicolumn{2}{c}{\textbf{Autoencoder - QCNN}} &
\multicolumn{2}{c}{\textbf{Noisy Simulation}}\\
\cline{3-6}
&&\textbf{Trace Distance}&\textbf{Accuracy (\%)}&
\textbf{Trace Distance} & 
\textbf{Accuracy (\%)} \\
\midrule
\multirow{2}{*}{\makecell{CIFAR-10\\(Airplane–Automobile)}} 
    & $g_a$ & $0.210 \pm 0.009$ &$65.1 \pm 9.7$  & $0.557 \pm 0.083$ & $82.60 \pm 4.08$\\
    & $g_c$ & $0.211 \pm 0.014$&$61.6 \pm 5.4$  & $0.460 \pm 0.065$ & $75.70 \pm 13.38$\\
\midrule
\multirow{2}{*}{\makecell{CIFAR-10\\(Frog–Ship)}} 
    & $g_a$ & $0.286 \pm 0.021$&$72.1 \pm 9.8$  & $0.811 \pm 0.025$ & $ 93.20 \pm 1.96$\\
    & $g_c$ & $0.414 \pm 0.034$&$78.8 \pm 6.1$  & $0.694 \pm 0.060$ & $91.0 \pm 1.70$\\
\midrule
\multirow{2}{*}{\makecell{Tiny ImageNet\\(School Bus-Maypole)}}
    & $g_a$ & $0.123 \pm 0.012$&$51.0 \pm 0.7$  & $0.767 \pm 0.031$ & $90.20 \pm 1.79$\\
    & $g_c$ & $0.225 \pm 0.016$&$62.8 \pm 7.4$  & $0.220 \pm 0.302$ & $64.80 \pm 17.71$\\
\bottomrule
\end{tabular}

\caption{Binary classification results of QCNNs with autoencoder-based embeddings on 4-qubit noiseless simulations, and the CNQE-based QCNN model under 4-qubit noisy simulations.
Each value represents the mean $\pm$ standard deviation over 5 independent runs.}
\label{tab: autoencoder_comparison_noisy_simulation}
\end{table*}

The results show that the CNQE–QCNN models reported in Table~\ref{tab: combined_results} consistently outperformed all autoencoder-based interface models, even when compared with the lowest-performing CNQE–QCNN configurations.
Moreover, relative to the classical baselines in Table~\ref{tab: classical_model_comparison}, the autoencoder models underperformed in every case when compared with the best-performing baseline model for each dataset. 

In summary, these findings indicate that CNQE provides a substantial improvement in classification accuracy on multi-channel datasets.

\subsection{Performance under Noisy Simulation}\label{subsec: Performance under Noisy Simulation}

To examine the robustness of CNQE-QCNN models under a realistic noisy quantum hardware setting, we conducted experiments using simulated quantum noise on a 4-qubit setup across all three datasets. For this analysis, we focused on fidelity-based loss and selected representative ansatz configurations for the $g_a$ and $g_c$ interface models. The results are summarized in Table~\ref{tab: autoencoder_comparison_noisy_simulation}, with detailed noise models and experimental settings provided in Appendix~\ref{Append: Experimental Setup for Noisy Simulation}.

Compared with the noiseless results in Table~\ref{tab: combined_results}, the $g_c$ model---which employs an embedding layer three times larger than that of $g_a$---exhibits greater sensitivity to noise across all datasets. This sensitivity was reflected in the trace distance: during CNQE training, $g_c$ was more strongly affected by noise, leading to a reduced trace distance. Consequently, QCNN training for $g_c$ experienced a larger performance drop than for $g_a$, due to the combined effect of reduced trace distance in CNQE training and noise acting within the QCNN stage.

When compared with the classical baselines in Table~\ref{tab: classical_model_comparison}, the noisy CNQE–QCNN models tended to show relatively robust performance overall. On the airplane–automobile dataset, the noisy CNQE–QCNN with $g_a$ outperformed the classical baseline, while the $g_c$ configuration fell behind. On the frog–ship dataset, CNQE–QCNN consistently outperformed the best classical model in both cases. For the school bus–maypole dataset, the noisy CNQE–QCNN achieved superior accuracy over the classical baseline in the $g_a$ configuration; however, the large variance in the $g_c$ configuration under noisy runs led to lower performance compared with the best classical model.

A key observation is that the noisy CNQE–QCNN models outperformed the autoencoder-based QCNN baselines across all settings, even though the latter were evaluated under noiseless conditions (Table~\ref{tab: autoencoder_comparison_noisy_simulation}). This is consistent with the role of trace distance in classification: under certain non-unitary CPTP maps, such as depolarizing channels, trace distance is strictly contractive and can only decrease. Embeddings that start with a larger trace distance therefore yields classifiers that are more robust to noise. CNQE explicitly optimizes for state distinguishability, whereas autoencoders compress inputs without controlling trace distance. As reported in Table~\ref{tab: autoencoder_comparison_noisy_simulation}, the trace distances obtained with the autoencoder were consistently smaller than those obtained with CNQE. Consequently, even in the presence of noise, classifiers using CNQE embeddings achieve higher accuracy than autoencoder-assisted classifiers evaluated in the noiseless setting.

Overall, CNQE–QCNN exhibits only modest accuracy degradation under noisy conditions, while still outperforming both classical baselines and autoencoder–QCNN models without noise. These results demonstrate that CNQE provides a noise-resilient embedding strategy with practical advantages for quantum machine learning on multi-channel data.

\section{Conclusions and Discussions}\label{sec: Conclusions and Discussions}

This study proposed three configurations of the CNQE model, each defined by a distinct interface model—$g_a$, $g_b$, and $g_c$—for the binary classification of multi-channel data. The proposed framework demonstrated that quantum embedding can be efficiently configured to achieve competitive performance.

The interface model $g_c$, which processes each channel separately and embeds each into a dedicated quantum layer, achieved a performance comparable to other models, while using the fewest trainable parameters.
This result is supported by the theoretical understanding that in $g_c$, where each channel is processed independently in the classical stage, interchannel information is nonlinearly formed within the quantum circuit after embedding. Consequently, quantum embedding effectively captures full-channel relationships without requiring explicit interchannel modeling in the classical network.
Consequently, $g_c$ leveraged the expressive power of the quantum circuit to achieve strong performance with minimal trainable parameters, highlighting the advantage of quantum-assisted feature representation.

Furthermore, we evaluated the performance of models constructed using various combinations of CNQE loss functions and embedding ansatzes. To assess the impact of these design choices, we performed statistical comparisons using Welch’s $t$-test. The results showed no statistically significant difference between the fidelity-based and Hilbert–Schmidt-based loss functions or among different ansatz types, indicating that the performance is largely robust to these design choices.

Although the trace distance between quantum states is mathematically bounded above by one, the values observed during CNQE training converged to dataset-dependent levels below this limit. 
This can be attributed to the simplicity of the classical neural network architecture used in the experiments. Using more expressive classical models could further increase the achievable trace distance, potentially improving the classification performance. Exploring the impact of classical model complexity on the trace distance and overall training dynamics constitutes an important direction for future research.

For real quantum devices, hybrid approaches that incorporate quantum error mitigation~\cite{endo2018practical,temme2017error,9226505,Kurita2023synergeticquantum,Maciejewski2020mitigationofreadout,9142431,kim2022quantum,PhysRevApplied.17.014024,Lee_2023} or adaptive models~\cite{AdaptVQE,QMI_semiagnostic,AdaptVQE-2,Wada_2024} could improve the performance of binary classification using the CNQE-based VQA model by reducing errors and enhancing robustness.

Beyond binary classification, NQE can be extended to a wider range of tasks that rely on the effective embedding of classical data into quantum circuits. The improved separability of quantum states in the Hilbert space can reduce the risk of barren plateaus~\cite{holmes_2022,hur2023neural}, which remains a major challenge in variational quantum algorithms.

\section*{Acknowledgments}
This work was supported by a Korea Research Institute for Defense Technology Planning and Advancement grant funded by Defense Acquisition Program Administration (DAPA) (KRIT-CT-23-031). This work was also supported by Institute of Information \& communications Technology Planning \& evaluation (IITP) grant funded by the Korea government (No. 2019-0-00003, Research and Development of Core technologies for Programming, Running, Implementing and Validating of Fault-Tolerant Quantum Computing System), the National Research Foundation of Korea (RS-2023-NR119931, RS-2025-02309510), and the Ministry of Trade, Industry, and Energy (MOTIE), Korea, under the Industrial Innovation Infrastructure Development Project (Project No. RS-2024-00466693).
\appendix
\section{Details of the Classical Model Used in CNQE Training} \label{Append: Details of the Classical Model Used in CNQE Training}
In this section, we present the architectural details of the interface models proposed in our study, as illustrated in Fig.~\ref{fig:CNN structures}. 
In addition, the trainable parameter counts are summarized in Table~\ref{tab: CNN parameters}.
Because our experiments are based on multi-channel image data with shape $(32 \times 32 \times 3)$, we describe the model architecture tailored to this input format.

Model $g_a$ takes an input that is reshaped to $3\times32\times32$ and processes it using two successive 2D convolutional layers.
Each layer doubles the number of channels using $3\times3$ kernels, followed by an ReLU activation~\cite{agarap2019deeplearningusingrectified} and a $4\times4$ max-pooling operation. 
This transforms the input into intermediate shapes of $6\times8\times8$ and $12\times2\times2$. A fully connected hidden layer with 48 nodes then generates an output vector of eight or ten dimensions, depending on the embedding circuit’s dimensionality.

Model $g_b$ shares the same CNN and first FC layer structure as model $g_a$ but produces a three times larger output vector, resulting in a 24- or 30-dimensional output.

In model $g_c$, each input channel is processed independently using separate weights. 
Each channel is passed through two successive 2D convolutional layers as in models $g_a$ and $g_b$. Each layer uses $3\times3$ kernels with channel doubling, followed by ReLU activation and $4\times4$ max pooling.
This yields feature maps progressing through shapes $1\times32\times32$, $2\times8\times8$, and $4\times2\times2$ for each channel. 
These are then flattened and passed through an FC hidden layer with 16 nodes, producing an eight- or ten-dimensional output per channel. The final output vector is obtained by concatenating the outputs from all channels, resulting in a 24- or 30-dimensional vector that matches the output dimensionality of model $g_b$.

In this specific setting, the model parameters are computed as follows:
To facilitate the comparison, we express the parameter calculations using variables.
Following the main text, $n_c$ denotes the number of input channels and $n_f$ denotes the output dimension of the CNN block (also the input dimension to the first FC layer). 
The final output vector dimension is represented by variable $p_a$, which refers to the dimension of the vector produced by a single quantum embedding layer. Accordingly, the final output vector dimensions for each model are $p_a$ for model $g_a$ and $n_c p_a$ for models $g_b$ and $g_c$.

In model $g_a$, the number of parameters in the two 2D convolutional layers is calculated as
\begin{align*}
\begin{split}
    \mathrm{First\,\,Conv2D}&: (n_c \times 2n_c \times 3 \times 3) + 2n_c = 18{n_c}^2+2n_c\\
    \mathrm{Second\,\,Conv2D}&: (2n_c \times 4n_c \times 3 \times 3) + 4n_c = 72{n_c}^2+4n_c
\end{split}    
\end{align*}
Each step calculation is performed using $(\mathrm{input\,\,channel}\times \mathrm{output\,\,channel}\times (\mathrm{kernel\,\,size}))+\mathrm{output\,\,channel}(\mathrm{bias})$, resulting in $90{n_c}^2+6n_c$ parameters for the CNN layers. The following parameters are calculated for the two FC layers:
\begin{align*}
\begin{split}
    \mathrm{First\,\,FC}&: (n_f \times n_f) + n_f = {n_f}^2 + n_f\\
    \mathrm{Second\,\,FC}&: (n_f \times p_a) + p_a = n_f p_a + p_a
\end{split}    
\end{align*}
where each step follows the formula: $(\mathrm{input\,\,feature}\times \mathrm{output\,\,feature})+\mathrm{output\,\,feature}(\mathrm{bias})$. The total number of parameters of the FC layers is ${n_f}^2 + n_f+ n_f p_a + p_a$.

In model $g_b$, 
The CNN structure is identical to that of model $g_a$. Therefore, the total number of CNN parameters is 
\(90{n_c}^2+6n_c\). 
 The parameters of the FC layers are calculated as follows: 
\begin{align*}
\begin{split}
    \mathrm{First\,\,FC}&: (n_f \times n_f) + n_f = {n_f}^2 + n_f\\
    \mathrm{Second\,\,FC}&: (n_f \times n_c p_a) + n_c p_a = n_c n_f p_a + n_c p_a
\end{split}    
\end{align*}
The total number of parameters of the FC layers is ${n_f}^2 + n_f+ n_c n_f p_a + n_c p_a$. 

Model $g_c$ consists of three separate CNN structures, where each channel includes two convolutional 2D layers. Their parameters are calculated as follows:
\begin{align*}
\begin{split}
    \mathrm{First\,\,Conv2D}&: \left(1 \times 2 \times 3 \times 3\right) + 2 = 20\\
    \mathrm{Second\,\,Conv2D}&: \left(2 \times 4 \times 3 \times 3\right) + 4 = 76
\end{split}    
\end{align*}
The total number of parameters for the CNN layers across all the three channels is $96\times n_c=96n_c$.
The FC layers have the following parameters:
\begin{align*}
\begin{split}
    \mathrm{First\,\,FC}&: \left(\frac{n_f}{n_c} \times \frac{n_f}{n_c}\right) + \frac{n_f}{n_c} = \frac{n_f^2}{n_c^2}+\frac{n_f}{n_c}\\
    \mathrm{Second\,\,FC}&: \left(\frac{n_f}{n_c} \times x\right) + p_a = \frac{n_f}{n_c}p_a+p_a
\end{split}    
\end{align*}
where the total number of parameters of the FC layers is $({n_f^2}/{n_c^2}+{n_f}/{n_c}+({n_f}/{n_c})p_a+p_a)\times n_c=
{n_f^2}/{n_c}+{n_f}+{n_f}p_a+n_c p_a$. 
Thus, the number of trainable parameters is in the order $g_b > g_c > g_a$.

\section{Fourier Analysis of Embedded Quantum State} \label{Append: Fourier Analysis of Embedded Quantum State}
In this section, we analyze the expressivity of the embedded quantum states on a Fourier basis, as shown in Eqs.~\eqref{eq: fourier analysis} and \eqref{eq: fourier analysis frequency}.

The quantum embedding circuit $U(\boldsymbol{x})$ is generally expressed as
\begin{equation}
U(\boldsymbol{x})=\left(\prod_{l=1}^{L}W_l\,U_l(\boldsymbol{x}_{l})\right)W_{L+1}
\end{equation}
where $W_l$ is a unparameterized unitary and $U_l$ is a data-encoding unitary for the $l$-th layer with input $\boldsymbol{x}_l$.
Without loss of generality, the data-encoding unitary $U_{l}(\boldsymbol{x}_{l})$ can be written as
\begin{align}
\begin{split}
U_{l}(\boldsymbol{x}_{l})
&=\prod_{q=1}^{n}\exp(-
\frac{i}{2} (\boldsymbol{x}_l)_q\sigma_z^{(q)}) \\
&=\exp\,\left(-\frac{i}{2}\sum^{n}_{q=1}
    (\boldsymbol{x}_l)_q\sigma_z^{(q)}\right)\\
    &=\exp\left(i D_l(\boldsymbol{x}_l)\right)
\end{split}
\end{align}
where $\sigma_z^{(q)}$ is the Pauli-$Z$ rotation gate acting on the $q$th qubit with the input, and $(\boldsymbol{x}_l)_q$ is the $q$th component of $\boldsymbol{x}_l$.
Here, $D_l(\boldsymbol{x}_l)\equiv(-1/2)\sum_{q=1}^{n}(\boldsymbol{x}_l)_q\,\sigma_z^{(q)}$.
In the first equality, we consider a universal gate set consisting of CNOT gates and all single-qubit rotation gates. For the second equality, we use the fact that single-qubit rotation gates can be decomposed into diagonal matrices involving $\sigma_z$. The remaining nondiagonal terms are absorbed into the neighboring unitary operators $W_l$ and $W_{l+1}$.
In the third equality, $D_l$ denotes a diagonal matrix with $2^n$ eigenvalues. Each eigenvalue corresponds to the sum of $\pm (\boldsymbol{x}_l)_q$ over $q = 1$ to $n$ up to a constant value.

The quantum state obtained after embedding the input $\boldsymbol{x}$ into the quantum circuit $U(\boldsymbol{x})$, starting from the initialized state $|\boldsymbol{0}_n\rangle$ is given by
\begin{align}
\begin{split}
    \left[U(\boldsymbol{x})|\boldsymbol{0}_n\rangle
    \right]_i=&\sum^{2^n}_{k_1,\cdots,k_L=1}
    (W_1)_{i\,k_1}U_1(\boldsymbol{x}_1)_{k_1k_1}
    (W_2)_{k_1k_2}\\
    &\qquad\qquad\cdots U_L(\boldsymbol{x}_L)_{k_Lk_L}(W_{L+1})_{k_L1}\\
    =&\sum_{\omega\in\Omega} c_{\omega}\exp(i\omega(\boldsymbol{x}))\,,
\end{split}
\end{align}
where the frequency components are defined as follows:
\begin{align}\label{eq: append_frequency}
\begin{split}
    \omega&\in\Omega=\{(D_1(\boldsymbol{x}_1))_{k_1k_1}+\cdots+
    (D_L(\boldsymbol{x}_L))_{k_Lk_L}\big|\\
    &\qquad\qquad\qquad\qquad k_i\in\{1,\cdots,2^n\}, 
    \forall i\in\{1,\cdots,L\} \}\,,
\end{split}
\end{align}
and the corresponding coefficients are
\begin{equation*}
    c_\omega=(W_1)_{i\,k_1}(W_2)_{k_1k_2}\cdots (W_{L+1})_{k_L1}\,.
\end{equation*}
Each frequency $\omega$ is characterized by the index combination $(k_1,\cdots,k_L)$ and expressed as the sum of the corresponding diagonal components. 
Because the diagonal components of each $D_l(\boldsymbol{x}_l)$ are linear combinations of the vector components of $\boldsymbol{x}_l$, the frequencies $\omega$ are linear combinations of all components across all input vectors for the $L$ layers. Thus, Eq.~\eqref{eq: append_frequency} can be rewritten as follows:
\begin{align}
\begin{split}
\omega\in\Omega
&=\left\{
    \sum^{n}_{i=1} {(\boldsymbol{h}_1)}_{i}(\boldsymbol{x}_1)_{i}
    +\cdots+
    {(\boldsymbol{h}_L)}_{i}(\boldsymbol{x}_L)_{i}
    \right\}\\
&=\left\{ \sum^{L}_{l=1} (\boldsymbol{h}_l)\cdot(\boldsymbol{x}_l)\right\}\\
&=\left\{ \boldsymbol{h}\cdot\boldsymbol{x} \right\}
\end{split}
\end{align}
where each \( \boldsymbol{h}_l \in \mathbb{R}^n \) for $l=\{1,\cdots,L\}$ denotes a possible coefficient vector in the linear combination that defines the frequency component \( \omega \).  
By defining the concatenated vectors \( \boldsymbol{x} = [\boldsymbol{x}_1, \dots, \boldsymbol{x}_L] \) and \( \boldsymbol{h} = [\boldsymbol{h}_1, \dots, \boldsymbol{h}_L] \), each element of \( \Omega \) can be compactly expressed as \( \boldsymbol{h} \cdot \boldsymbol{x} \).
Using this framework, we derive Eqs.~\eqref{eq: fourier analysis} and \eqref{eq: fourier analysis frequency}.

\section{Experimental Setup: Data and Training Details} \label{Append: Experimental Setup: Data and Training Details}

\begin{figure}[h]
\centering
\includegraphics[width=0.95\columnwidth]{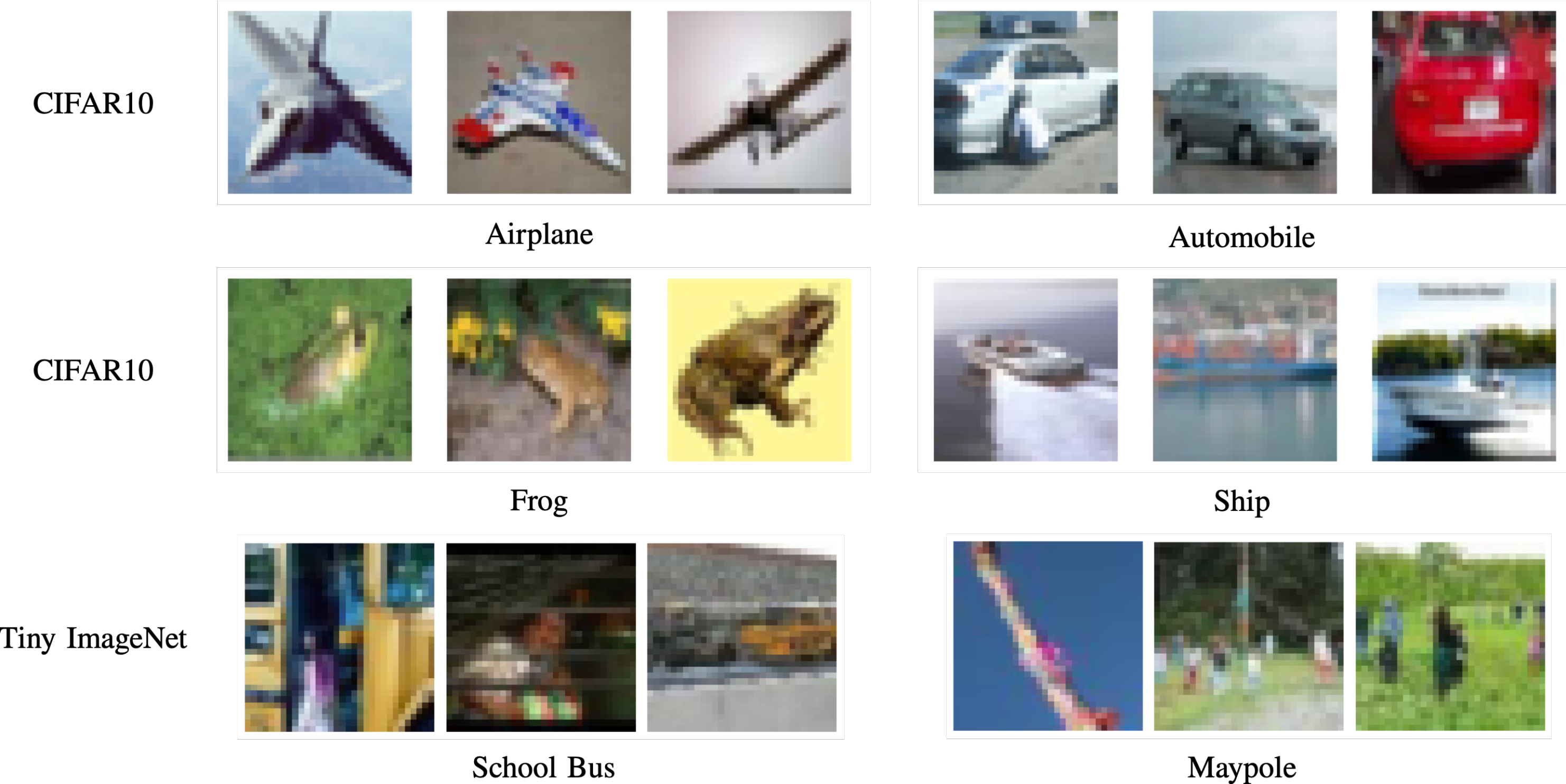}
\caption{\label{fig: Data} Datasets used in this study.}
\end{figure}

CIFAR-10 consists of ten distinct classes of images of size $32\times32$ in three color channels (RGB). Tiny ImageNet comprises 200 classes with images of size $64\times64$ in three color channels (RGB). Three pairs of classes are selected for evaluating the binary classification performance: Airplane–Automobile and Frog–Ship from CIFAR-10, and School Bus–Maypole from Tiny ImageNet. Representative examples of these pairs are presented in Fig.~\ref{fig: Data}.

All data are resized to $3\times32\times32$, and the pixel values are scaled to the range $[0, 1]$ without further normalization. For Tiny ImageNet, downsampling is performed via interpolation between pixels.

A total of 800 samples (400 per class) are used for training and 200 samples are allocated for testing. During CNQE training, a batch of 25 sample pairs is randomly drawn from 800 training examples at each iteration. Initially, all models are trained for 2000 iterations. The optimal number of iterations is then selected by identifying the point at which the trace distance on the test data saturates or begins to decrease. Training is conducted for the selected number of iterations and repeated over five independent runs. The final CNQE parameters are selected based on the median trace distance of the test data across five runs.

\begin{figure}[t]
\centering
\includegraphics[width=0.95\columnwidth]{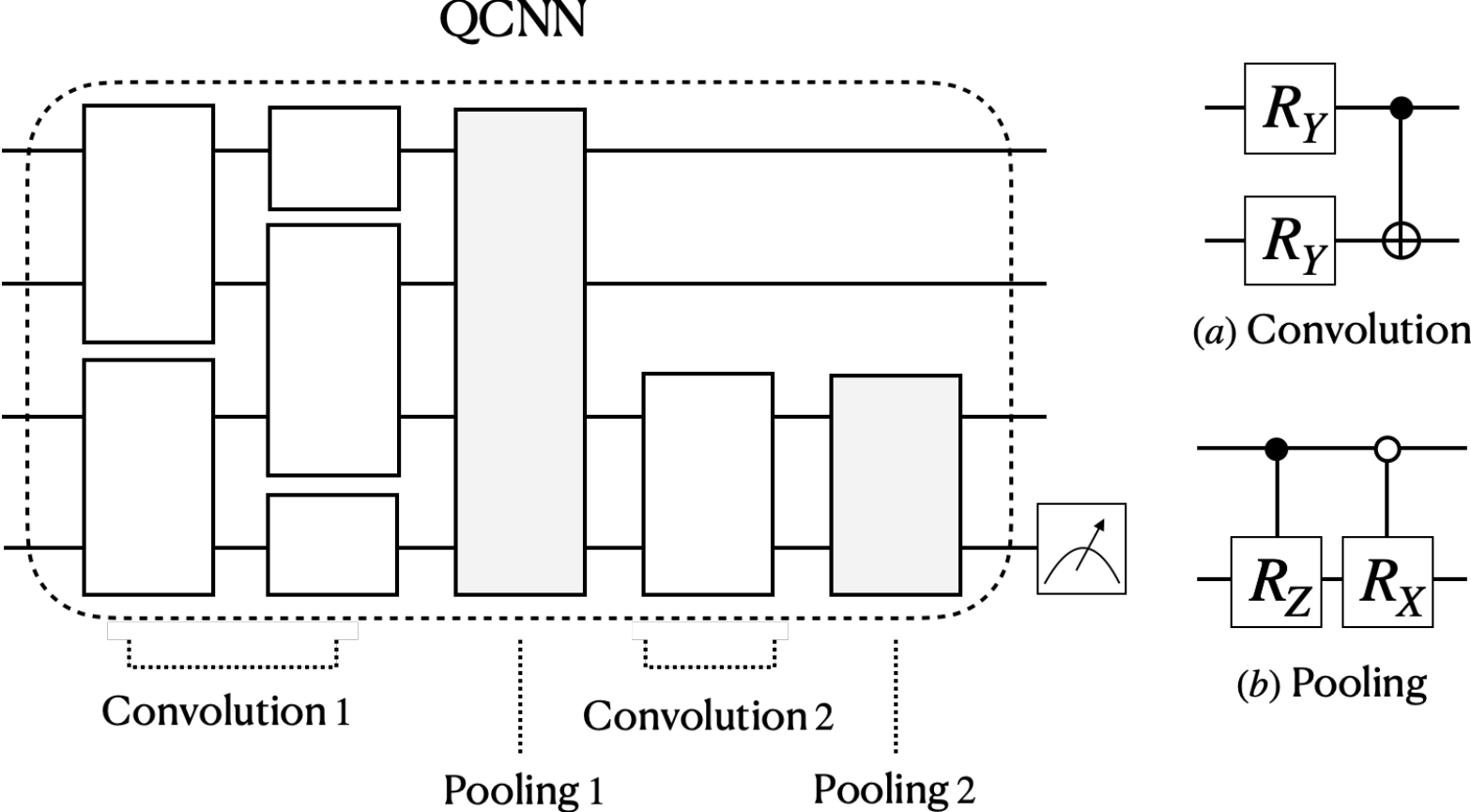}
\caption{\label{fig: QCNN}  QCNN configuration for Ansatz $V(\theta)$ in VQA training, as illustrated in Fig.~\ref{fig:Figure1}. QCNN consists of two convolutional and two pooling layers, each implemented with unit circuits $(a)$ and $(b)$.}
\end{figure}

Following CNQE optimization, we perform the ansatz training. The variational quantum circuit $V(\theta)$ used in the VQA model (Fig.~\ref{fig:Figure1}), we adopt a quantum convolutional neural network (QCNN)~\cite{cong_quantum_2019}.
To evaluate the effect of data embeddings independent of the circuit complexity, we employ the simplest unit circuits introduced in~\cite{hur2022quantum}, as illustrated in Fig.~\ref{fig: QCNN}.

QCNN is trained using a batch size of 5 over 800 training samples for 20 epochs, repeated across five independent runs. The model performance is evaluated using the test data.
All training is performed using the Adam optimizer~\cite{kingma2017adam} with a learning rate of 0.01, as implemented in Qiskit~\cite{Qiskit} and PennyLane~\cite{bergholm2020pennylane}.

\section{Extended Experiments with 6- and 8-Qubit Systems} \label{Append: Extended Experiments with 8-Qubit Simulation}

\begin{table*}[!]
    \centering
    \setlength{\tabcolsep}{5pt}
    \renewcommand{\arraystretch}{1.1}
    \begin{tabular}{c}
        \begin{tabular}{c|c|c|c|c|c}
            \toprule
            \textbf{Qubit System} &
            \textbf{Data} &
            \textbf{\makecell{Interface \\ Model}} &
            \textbf{\makecell{Embedding \\ Circuit}} & \textbf{\makecell{Trace \\ Distance}} & \textbf{Accuracy (\%)} \vspace{0.1cm}\\
            \hline
            \multirow{6}{*}{6 qubit} 
            &\multirow{3}{*}{\makecell{CIFAR-10\\(Frog–Ship)}} 
            &$g_a$ & $(b)$ & $0.840 \pm 0.03$ & $94.3 \pm 1.60$ \\
            &&$g_b$ & $U_{\mathrm{NC}}$ & $0.842 \pm 0.07$ & $94.6 \pm 1.39$ \\
            &&$g_c$ & $U_{\mathrm{NC}}$ & $0.806 \pm 0.03$ & $93.4 \pm 1.98$ \\
            
            \cmidrule{2-6}
            &\multirow{3}{*}{\makecell{Tiny ImageNet\\(School Bus-Maypole)}} 
            &$g_a$ & $(b)$ & $0.842 \pm 0.03$ & $93.3 \pm 2.46$ \\
            &&$g_b$ & $U_{\mathrm{NC}}$ & $0.843 \pm 0.02$ & $93.7 \pm 1.40$ \\
            &&$g_c$ & $U_{\mathrm{NC}}$ & $0.696 \pm 0.09$ & $89.3 \pm 4.84$ \\    
            
            \midrule
            \multirow{6}{*}{8 qubit} 
            &\multirow{3}{*}{\makecell{CIFAR-10\\(Frog–Ship)}} 
            &$g_a$ & $(b)$ & $0.849 \pm 0.02$ & $94.8 \pm 1.15$ \\
            &&$g_b$ & $U_{\mathrm{NC}}$ & $0.841 \pm 0.02$ & $94.8 \pm 1.04$ \\
            &&$g_c$ & $U_{\mathrm{NC}}$ & $0.731 \pm 0.09$ & $92.3 \pm 1.48$ \\
            
            \cmidrule{2-6}
            &\multirow{3}{*}{\makecell{Tiny ImageNet\\(School Bus-Maypole)}} 
            &$g_a$ & $(b)$ & $0.819 \pm 0.06$ & $91.0 \pm 3.08$ \\
            &&$g_b$ & $U_{\mathrm{NC}}$ & $0.834 \pm 0.05$ & $91.6 \pm 3.36$ \\
            &&$g_c$ & $U_{\mathrm{NC}}$ & $0.741 \pm 0.05$ & $90.4 \pm 4.22$ \\
            \bottomrule
        \end{tabular}
    \end{tabular}

    \caption{
        Performance comparisons among interface models 
        on the CIFAR-10 frog–ship and Tiny ImageNet school bus-maypole class pairs under 6- and 8-qubit settings.
        Each value is reported as the mean $\pm$ standard deviation across five independent runs.
    }
    \label{tab: results_68qubit}
\end{table*}

\begin{figure*}[t]
\centering
\includegraphics[width=0.9\textwidth]{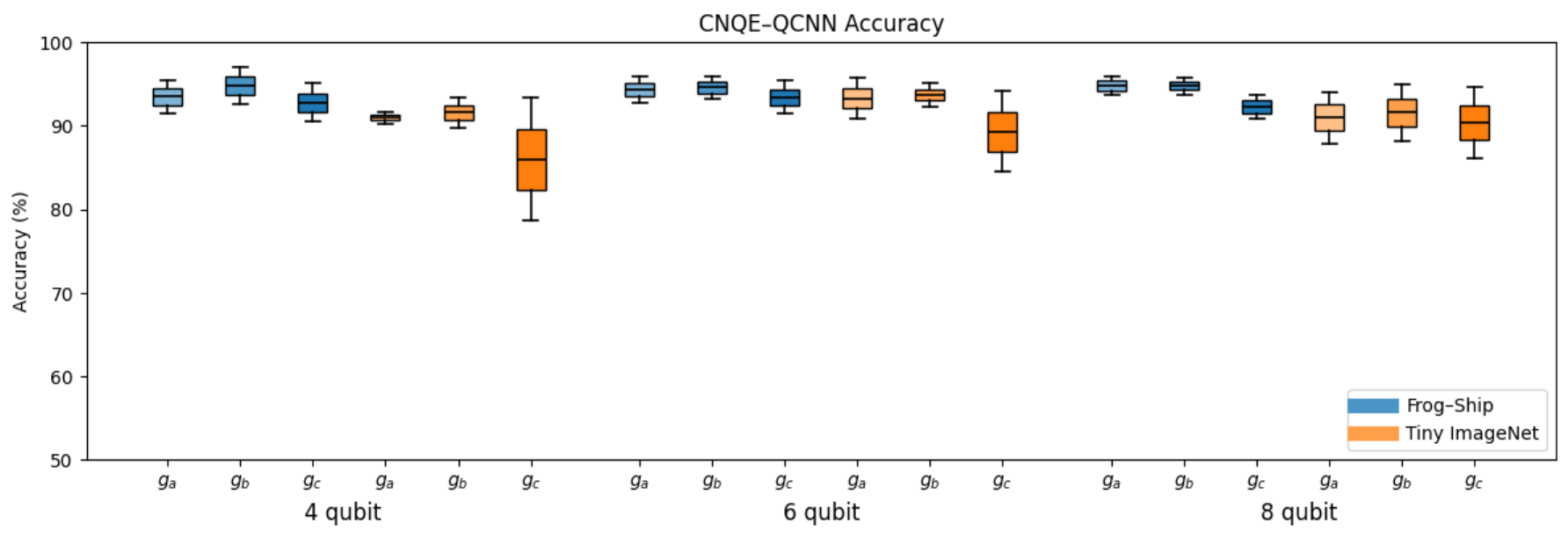}
\caption{\label{fig: boxplots_468qubit}Binary classification accuracy of CNQE-QCNN on the CIFAR-10 frog–ship (blue) and Tiny ImageNet school bus-maypole (orange) datasets for 4-, 6-, 8-qubit systems. Results are shown for three interface models ($g_a, g_b, g_c$). Each box summarizes the distribution of accuracy over five runs.}
\end{figure*}

To assess scalability, we extended the experiments beyond the 4-qubit setting and evaluated selected configurations on 6- and 8-qubit systems. This analysis focuses on how performance is affected by increasing system size. A representative subset of models from the 4-qubit case was analyzed: $g_a$ with ansatz $(b)$ and $g_b$ and $g_c$ with the $U_{\mathrm{NC}}$ ansatz, all trained using the Hilbert–Schmidt loss.

The classical part of each interface model follows a 4-qubit design, with the final FC layer’s output dimension 
scaled proportionally to the number of qubits.
Specifically, for a 6-qubit system, the output dimension of $g_a$ is set to 12 and $g_b$ to 36, while each channel output of $g_c$ is set to 12, resulting in a total output vector of size 36. For an 8-qubit system, these values increase to 16, 48, and 16 per channel (total 48), respectively.
This scaling is consistent with our analysis
in Section~\ref{subsec: Configurations: Embedding Circuit}, where we showed that the number of embedding parameters in both ansatz $(b)$ and $U_{\mathrm{NC}}$ 
grows proportionally with the qubit count.

All models were on the CIFAR-10 frog–ship class and the Tiny ImageNet school bus–maypole class pairs.
Each experiment was repeated five times with random initialization. The hyperparameters are identical to those used in the 4-qubit setup.

The results are summarized in Table~\ref{tab: results_68qubit}, and Figure~\ref{fig: boxplots_468qubit} presents accuracy distributions across 4-, 6-, and 8-qubit systems. The findings show that increasing the number of qubits does not degrade performance; accuracy remains stable with only minor variations. These results demonstrate that CNQE-QCNN maintains robust performance as the qubit count increases, consistently outperforming the classical baseline.

\section{Trace Distance Statistics by Model Component} \label{Append: Trace Distance Statistics by Model Component}

Table~\ref{tab:trace_distance_summary} summarizes the results discussed in Section~\ref{subsec: Numerical Experiments: Effect of Model Architecture on Trace Distance}.

\newcolumntype{L}[1]{>{\raggedright\arraybackslash}p{#1}} 
\newcolumntype{C}[1]{>{\centering\arraybackslash}p{#1}}   

\begin{table}[!]
    \centering
    \renewcommand{\arraystretch}{1.2}
    \begin{tabular}{C{2.6cm} C{2.5cm} C{2.6cm}} 
        \toprule
        \textbf{Category} & \textbf{Component} & \textbf{Trace Distance} \\
        \midrule
        \multirow{3}{*}{Interface Model} 
            & $g_a$ & $0.742 \pm 0.131$ \\
            & $g_b$ & $0.746 \pm 0.141$ \\
            & $g_c$ & $0.652 \pm 0.113$ \\
        \midrule
        \multirow{2}{*}{Loss Function} 
            & Fidelity & $0.720 \pm 0.118$ \\
            & Hilbert-Schmidt & $0.718 \pm 0.150$ \\
        \midrule
        \multirow{7}{*}{\makecell{Embedding Circuit\\under $g_a$}}
            & (a) & $0.736 \pm 0.163$ \\
            & (b) & $0.762 \pm 0.097$ \\
            & (c) & $0.748 \pm 0.103$ \\
            & (d) & $0.762 \pm 0.097$ \\
            & (e) & $0.750 \pm 0.114$ \\
            & (f) & $0.700 \pm 0.184$ \\
            & (g) & $0.734 \pm 0.130$ \\
        \midrule
        \multirow{4}{*}{\makecell{Embedding Circuit\\under $g_b$ and $g_c$}}
            & $U_{\mathrm{ZZ}}$ & $0.692 \pm 0.133$ \\
            & $U_{\mathrm{NC}}$ & $0.711 \pm 0.119$ \\
            & $U_{\mathrm{NC10}}$ & $0.711 \pm 0.124$ \\
            & $U_{\mathrm{NCL}}$ & $0.681 \pm 0.163$ \\
        \bottomrule
    \end{tabular}
    \caption{Trace distance statistics across interface models, loss functions, and embedding circuits. Each value represents the mean $\pm$ standard deviation computed from five independent runs per configuration.}
    \label{tab:trace_distance_summary}
\end{table}

\section{Summary of Classical Baseline Models} \label{Append: Summary of Classical Baseline Model Performance}
 
Table~\ref{tab: classical_model_comparison} compares classical model architectures and their classification performance, as discussed in Section ~\ref{subsec: Performance Comparison with Classical Baseline Models}.

\begin{table*}[!]
\centering

\begin{minipage}{\linewidth}
\centering
\renewcommand{\arraystretch}{1.5}
\begin{tabular}  
{
  >{\centering\arraybackslash}m{2cm}| 
  >{\centering\arraybackslash}m{2cm}| 
  >{\centering\arraybackslash}m{0.5cm}
 >{\centering\arraybackslash}m{6cm} |
  >{\centering\arraybackslash}m{2cm}
}
\toprule
  \textbf{\makecell{Interface \\ Model}}& \textbf{\makecell{Final output \\ dimensionality \\ of CNQE}} & &\textbf{Classical Baseline Model} & \textbf{\makecell{\# of trainable \\ Parameters}}\\
\midrule
\multirow{3}{*}{$g_a$} 
& \multirow{3}{*}{8} 
& (1)& Linear(8,2) & 18 \\\cline{3-5}
& & (2)& Linear(8,1)$\rightarrow$Relu$\rightarrow$Linear(1,2) & 13 \\\cline{3-5}
& & (3)& CNN1D(kernel=6)$\rightarrow$Relu$\rightarrow$Linear(3,2) & 15 \\
\midrule
\multirow{3}{*}{$g_b$ or $g_c$} 
& \multirow{3}{*}{24} 
& (1)& Linear(24,2) & 50 \\\cline{3-5}
& & (2)& Linear(24,1)$\rightarrow$Relu$\rightarrow$Linear(1,2) & 29 \\\cline{3-5}
& & (3)& \vspace{3pt}\makecell{CNN1D(kernel=3)$\rightarrow$Relu$\rightarrow$ \\
CNN1D(kernel=2)$\rightarrow$Relu$\rightarrow$ \\
CNN1D(kernel=2)$\rightarrow$Relu$\rightarrow$Linear(3,2)} & 18 \\
\bottomrule
\end{tabular}
\vspace{2pt}
\caption*{\small (a) }
\end{minipage}

\vspace{1.2em}  
\begin{minipage}{\linewidth}
\centering
\setlength{\tabcolsep}{10pt}
\renewcommand{\arraystretch}{1.2}
\begin{tabular}{c|c|c|c}
    \toprule
    \textbf{Data} & \textbf{Interface Model} & \textbf{Classical Baseline Model} & \textbf{Accuracy (\%)} \\
    \midrule
    \multirow{9}{*}{\makecell{CIFAR-10\\(Airplane–Automobile)}} & \multirow{3}{*}{$g_a$} & (1) & \textbf{71.5} $\pm$ 19.7 \\
                                   &      & (2) & 57.8 $\pm$ 17.4 \\
                                   &      & (3) & 63.0 $\pm$ 17.8 \\
    \cmidrule{2-4}
                                   & \multirow{3}{*}{$g_b$} & (1) & \textbf{77.5} $\pm$ 15.6 \\
                                   &       & (2) & 50.0 $\pm$ 0.0 \\
                                   &       & (3) & 56.2 $\pm$ 13.9 \\
    \cmidrule{2-4}
                                   & \multirow{3}{*}{$g_c$} & (1) & \textbf{81.3} $\pm$ 2.1 \\
                                   &       & (2) & 50.0 $\pm$ 0.0 \\
                                   &       & (3) & 55.9 $\pm$ 13.2 \\
    \midrule
    \multirow{9}{*}{\makecell{CIFAR-10\\(Frog–Ship)}}          & \multirow{3}{*}{$g_a$} & (1) & \textbf{92.6} $\pm$ 1.7 \\
                                  &      & (2) & 58.7 $\pm$ 19.5 \\
                                  &      & (3) & 75.4 $\pm$ 23.2 \\
    \cmidrule{2-4}
                                  & \multirow{3}{*}{$g_b$} & (1) & \textbf{90.8} $\pm$ 3.2 \\
                                  &       & (2) & 50.0 $\pm$ 0.0 \\
                                  &       & (3) & 59.4 $\pm$ 21.0 \\
    \cmidrule{2-4}
                                  & \multirow{3}{*}{$g_c$} & (1) & \textbf{91.9} $\pm$ 1.7 \\
                                  &       & (2) & 50.0 $\pm$ 0.0 \\
                                  &       & (3) & 50.0 $\pm$ 0.0 \\
    \midrule
    \multirow{9}{*}{\makecell{Tiny ImageNet\\(School Bus-Maypole)}} & \multirow{3}{*}{$g_a$} & (1) & \textbf{86.4} $\pm$ 11.6 \\
                                       &      & (2) & 50.0 $\pm$ 0.0 \\
                                       &      & (3) & 77.6 $\pm$ 25.2 \\
    \cmidrule{2-4}
                                       & \multirow{3}{*}{$g_b$} & (1) & \textbf{93.2} $\pm$ 0.8 \\
                                       &       & (2) & 50.0 $\pm$ 0.0 \\
                                       &       & (3) & 50.0 $\pm$ 0.0 \\
    \cmidrule{2-4}
                                       & \multirow{3}{*}{$g_c$} & (1) & \textbf{84.0} $\pm$ 2.6 \\
                                       &       & (2) & 50.0 $\pm$ 0.0 \\
                                       &       & (3) & 50.0 $\pm$ 0.0 \\
    \bottomrule
\end{tabular}
\vspace{2pt}
\caption*{\small (b)}
\end{minipage}
\caption{
Comparison of classical model architectures and their classification performance. (a) Classical model architectures and their trainable parameters. (b) Mean and standard deviation of classification accuracy for each model on three datasets. The best result for each dataset–interface combination is highlighted in bold.
}
\label{tab: classical_model_comparison}
\end{table*}

\section{Precision, Recall, and F1-Scores for Binary Classification} \label{Append: Precision, Recall, and F1-Score for Binary Classification Experiments}

In the main text, accuracy served as the primary performance metric. Here we additionally report precision, recall, and F1-score, which provide complementary perspectives: precision reflects the reliability of positive predictions, recall measures the ability to identify relevant instances, and F1-score offers a balanced summary of the two. By including these measures, we provide a more comprehensive understanding of the binary classification results.

These metrics were evaluated for the binary classification tasks across all configurations previously reported in Table~\ref{tab: combined_results}, and for the classical baseline experiments reported in Table~\ref{tab: classical_model_comparison}. The corresponding results are presented in Table~\ref{tab:combined_results_prf} and Table~\ref{tab: classical_model_comparison_prf}. These supplementary results provide additional evidence and offer a more nuanced perspective on the performance characteristics of the proposed CNQE-QCNN models.

We first examine the $g_c$ models. For each dataset, we compared the CNQE-QCNN model with the highest accuracy to the best-performing classical baseline (both highlighted in bold in the tables). Across all datasets, CNQE-QCNN achieved higher precision, recall, and F1-scores than the classical baselines.
For the $g_a$ models, the comparison showed that on CIFAR-10 automobile–airplane, CNQE-QCNN attained higher precision and F1-scores but slightly lower recall compared to the classical baseline. On the other datasets (CIFAR-10 frog–ship and Tiny ImageNet), CNQE-QCNN outperformed the baselines in all three metrics.
For the $g_b$ models, CNQE-QCNN again exhibited higher precision and F1-scores but lower recall on CIFAR-10 automobile–airplane, while outperforming the baselines in all metrics on CIFAR-10 frog–ship. On Tiny ImageNet, CNQE-QCNN achieved higher precision but lower recall and F1-scores than the classical baseline.

Overall, CNQE-QCNN models generally outperform classical baselines, with improvements observed across most datasets and interface models. For the $g_a$ and $g_b$ configurations, CNQE-QCNN often achieved higher precision and F1-scores, even when recall was sometimes lower, suggesting a tendency to favor more reliable positive predictions while maintaining competitive overall performance. Notably, the $g_c$ configuration exhibited the most consistent and robust improvements, surpassing the classical baselines across all datasets and evaluation metrics. This consistency highlights $g_c$ as the configuration in which the advantages of CNQE-QCNN are most evident.
Moreover, for both $g_b$ and $g_c$, the classical components replacing the QCNN contained more than three times as many parameters, further underscoring the efficiency and competitiveness of the proposed approach.

\newcolumntype{M}[1]{>{\centering\arraybackslash}m{#1}}  
\newcommand{\NA}{---} 

\begin{table*}[!]
    \centering
    \begin{tabular}{M{3cm} | M{1.3cm} M{1.8cm} | M{1.8cm} M{1.8cm} M{1.8cm} | M{1.8cm} M{1.8cm} M{1.8cm}}
        \toprule
        \multirow{2}{*}{\textbf{Data}} &
        \multirow{2}{*}{\makecell{\textbf{Interface} \\ \textbf{Model}}} &
        \multirow{2}{*}{\makecell{\textbf{Embedding} \\ \textbf{Circuit}}} &
        \multicolumn{3}{c|}{\textbf{Fidelity-based}} &
        \multicolumn{3}{c}{\textbf{Hilbert-Schmidt-based}} \\
        \cmidrule{4-9}
        & & &
        \makecell{\textbf{Precision}\\(\%)} &
        \makecell{\textbf{Recall}\\(\%)} &
        \makecell{\textbf{F1-score}\\(\%)} &
        \makecell{\textbf{Precision}\\(\%)} &
        \makecell{\textbf{Recall}\\(\%)} &
        \makecell{\textbf{F1-score}\\(\%)} \\
        \bottomrule

        \vspace{0.1cm}
        \multirow{15}{*}{\makecell{CIFAR-10\\(Airplane–Automobile)}}
        &\multirow{7}{*}{$g_a$}
        & $(a)$ & $83.07 \pm 0.08$ & $83.4 \pm 0.49$ & $83.23 \pm 0.29$ & $77.81 \pm 0.69$ & $90.4 \pm 0.49$ & $83.63 \pm 0.27$ \\
        && $(b)$ & $81.23 \pm 0.69$ & $86.4 \pm 3.20$ & $83.69 \pm 1.25$ & $\mathbf{85.86 \pm 1.04}$ & $\mathbf{89.6 \pm 1.74}$ & $\mathbf{87.67 \pm 0.32}$ \\
        && $(c)$ & $80.36 \pm 0.49$ & $89.2 \pm 0.40$ & $84.55 \pm 0.18$ & $85.26 \pm 0.06$ & $86.8 \pm 0.40$ & $86.03 \pm 0.23$ \\
        && $(d)$ & $75.08 \pm 0.28$ & $92.8 \pm 0.40$ & $83.01 \pm 0.29$ & $82.07 \pm 0.51$ & $87.8 \pm 1.47$ & $84.83 \pm 0.42$ \\
        && $(e)$ & $79.82 \pm 0.00$ & $87.0 \pm 0.00$ & $83.25 \pm 0.00$ & $82.65 \pm 0.50$ & $83.8 \pm 0.98$ & $83.22 \pm 0.29$ \\
        && $(f)$ & $84.30 \pm 0.47$ & $78.4 \pm 1.50$ & $81.24 \pm 0.90$ &  $79.99 \pm 1.10$ & $84.6 \pm 0.80$ & $82.22 \pm 0.18$ \\
        && $(g)$ & $77.25 \pm 0.23$ & $94.4 \pm 0.49$ & $84.97 \pm 0.24$ & $89.00 \pm 0.44$ & $77.6 \pm 0.49$ & $82.91 \pm 0.09$ \\
        \cmidrule{2-9}
        &\multirow{4}{*}{$g_b$}
        & $U_{\mathrm{ZZ}}$   & $\mathbf{84.13 \pm 0.47}$ & $\mathbf{88.0 \pm 0.89}$ & $\mathbf{86.02 \pm 0.62}$ & $76.62 \pm 2.15$ & $84.2 \pm 0.40$ & $80.21 \pm 1.04$ \\
        && $U_{\mathrm{NC}}$  & $79.82 \pm 0.00$ & $91.0 \pm 0.00$ & $85.05 \pm 0.00$ & $87.64 \pm 0.44$ & $80.8 \pm 0.40$ & $84.08 \pm 0.25$ \\
        && $U_{\mathrm{NC10}}$& $79.41 \pm 0.56$ & $91.0 \pm 0.00$ & $84.81 \pm 0.32$ & $80.22 \pm 3.28$ & $87.2 \pm 3.25$ & $83.46 \pm 1.29$ \\
        && $U_{\mathrm{NCL}}$ & $83.06 \pm 0.15$ & $69.6 \pm 1.20$ & $75.73 \pm 0.64$ & $76.28 \pm 0.36$ & $86.8 \pm 0.40$ & $81.20 \pm 0.15$ \\
        \cmidrule{2-9}
        &\multirow{4}{*}{$g_c$}
        & $U_{\mathrm{ZZ}}$   & $80.33 \pm 3.13$ & $88.6 \pm 2.94$ & $84.15 \pm 0.35$ & $80.86 \pm 2.00$ & $93.4 \pm 0.80$ & $86.66 \pm 1.00$ \\
        && $U_{\mathrm{NC}}$  & $72.88 \pm 3.54$ & $95.6 \pm 1.96$ & $82.61 \pm 1.57$ & $77.94 \pm 1.51$ & $82.6 \pm 3.38$ & $80.17 \pm 1.96$ \\
        && $U_{\mathrm{NC10}}$& $\mathbf{85.24 \pm 0.12}$ & $\mathbf{86.6 \pm 0.80}$ & $\mathbf{85.91 \pm 0.46}$ & $74.94 \pm 2.71$ & $83.0 \pm 2.10$ & $78.69 \pm 0.61$ \\
        && $U_{\mathrm{NCL}}$ & $79.63 \pm 2.74$ & $83.8 \pm 1.60$ & $81.61 \pm 1.12$ & $82.69 \pm 0.85$ & $83.0 \pm 1.10$ & $82.83 \pm 0.28$ \\
        \bottomrule

        \vspace{0.1cm}
        \multirow{15}{*}{\makecell{CIFAR-10\\(Frog–Ship)}}
        &\multirow{7}{*}{$g_a$}
        & $(a)$ & $\mathbf{94.05 \pm 0.87}$ & $\mathbf{97.8 \pm 0.40}$ & $\mathbf{95.88 \pm 0.37}$ & $94.57 \pm 0.75$ & $93.8 \pm 0.40$ & $94.18 \pm 0.22$ \\
        && $(b)$ & $96.85 \pm 0.03$ & $92.2 \pm 0.98$ & $94.46 \pm 0.53$ & $76.75 \pm 4.25$ & $99.6 \pm 0.80$  & $86.61 \pm 2.29$ \\
        && $(c)$ & $91.35 \pm 0.00$ & $95.0 \pm 0.00$ & $93.14 \pm 0.00$ & $93.50 \pm 3.15$ & $92.0 \pm 9.21$  & $92.32 \pm 3.89$ \\
        && $(d)$ & $96.66 \pm 1.79$ & $93.8 \pm 4.40$ & $95.18 \pm 3.11$ & $90.68 \pm 0.34$ & $93.4 \pm 1.36$  & $92.02 \pm 0.55$ \\
        && $(e)$ & $90.74 \pm 0.00$ & $98.0 \pm 0.00$ & $94.23 \pm 0.00$ & $91.15 \pm 0.82$ & $94.6 \pm 0.49$  & $92.84 \pm 0.19$ \\
        && $(f)$ & $93.61 \pm 1.01$ & $96.4 \pm 0.80$ & $94.98 \pm 0.15$ & $89.62 \pm 0.00$ & $95.0 \pm 0.00$  & $92.23 \pm 0.00$ \\
        && $(g)$ & $94.00 \pm 0.00$ & $94.0 \pm 0.00$ & $94.00 \pm 0.00$ & $86.75 \pm 0.32$ & $93.0 \pm 0.00$  & $89.77 \pm 0.17$ \\
        \cmidrule{2-9}
        &\multirow{4}{*}{$g_b$}
        & $U_{\mathrm{ZZ}}$   & $95.95 \pm 0.85$ & $94.6 \pm 0.49$ & $95.27 \pm 0.22$ & $86.14 \pm 17.25$ & $90.4 \pm 7.42$  & $87.78 \pm 13.13$ \\
        && $U_{\mathrm{NC}}$  & $87.98 \pm 0.63$ & $98.0 \pm 0.00$ & $92.72 \pm 0.35$ & $\mathbf{96.04 \pm 0.00}$ & $\mathbf{97.0 \pm 0.00}$  & $\mathbf{96.52 \pm 0.00}$ \\
        && $U_{\mathrm{NC10}}$& $95.92 \pm 0.00$ & $94.0 \pm 0.00$ & $94.95 \pm 0.00$ & $86.63 \pm 3.32$ & $98.8 \pm 0.40$  & $92.29 \pm 2.08$ \\
        && $U_{\mathrm{NCL}}$ & $95.88 \pm 0.00$ & $93.0 \pm 0.00$ & $94.42 \pm 0.00$ & $87.20 \pm 18.21$ & $97.4 \pm 0.80$  & $90.91 \pm 12.24$ \\
        \cmidrule{2-9}
        &\multirow{4}{*}{$g_c$}
        & $U_{\mathrm{ZZ}}$   & $94.01 \pm 0.89$ & $90.8 \pm 0.98$ & $92.37 \pm 0.08$ & $86.27 \pm 18.83$ & $79.8 \pm 23.09$ & $78.48 \pm 17.88$ \\
        && $U_{\mathrm{NC}}$  & $\mathbf{93.82 \pm 0.72}$ & $\mathbf{97.0 \pm 0.00}$ & $\mathbf{95.38 \pm 0.37}$ & $90.32 \pm 0.67$ & $97.0 \pm 0.00$  & $93.54 \pm 0.36$ \\
        && $U_{\mathrm{NC10}}$& $93.53 \pm 1.05$ & $95.2 \pm 0.40$ & $94.35 \pm 0.35$ & $75.70 \pm 32.03$ & $74.2 \pm 33.61$ & $74.82 \pm 33.04$ \\
        && $U_{\mathrm{NCL}}$ & $91.25 \pm 1.01$ & $97.8 \pm 0.40$ & $94.40 \pm 0.35$ & $85.43 \pm 21.97$ & $83.4 \pm 6.77$  & $83.40 \pm 15.55$ \\
        \bottomrule

        \vspace{0.1cm}
        \multirow{15}{*}{\makecell{Tiny ImageNet\\(School Bus–Maypole)}}
        &\multirow{7}{*}{$g_a$}
        & $(a)$ & $\mathbf{97.83 \pm 0.00}$ & $\mathbf{90.0 \pm 0.00}$  & $\mathbf{93.75 \pm 0.00}$ & $95.72 \pm 0.14$ & $89.6 \pm 2.94$  & $92.54 \pm 1.66$ \\
        && $(b)$ & $93.98 \pm 0.74$ & $87.2 \pm 1.60$  & $90.45 \pm 0.54$ & $93.96 \pm 4.05$ & $74.8 \pm 12.81$ & $82.53 \pm 7.27$ \\
        && $(c)$ & $97.83 \pm 0.00$ & $90.0 \pm 0.00$  & $93.75 \pm 0.00$  & $90.49 \pm 1.65$ & $90.8 \pm 0.98$  & $90.62 \pm 0.35$ \\
        && $(d)$ & $100.00 \pm 0.00$ & $80.4 \pm 0.80$  & $89.13 \pm 0.49$ & $94.16 \pm 3.07$ & $81.6 \pm 2.33$  & $87.37 \pm 1.20$ \\
        && $(e)$ & $93.72 \pm 0.05$ & $89.6 \pm 0.80$  & $91.61 \pm 0.45$ & $97.26 \pm 0.85$ & $84.8 \pm 0.98$  & $90.60 \pm 0.45$ \\
        && $(f)$ & $93.93 \pm 0.06$ & $92.8 \pm 0.98$  & $93.36 \pm 0.52$ & $98.59 \pm 1.15$ & $82.0 \pm 1.79$  & $89.51 \pm 0.67$ \\
        && $(g)$ & $95.65 \pm 0.00$ & $88.0 \pm 0.00$  & $91.67 \pm 0.00$ & $97.54 \pm 0.03$ & $79.2 \pm 0.98$  & $87.41 \pm 0.61$ \\
        \cmidrule{2-9}
        &\multirow{4}{*}{$g_b$}
        & $U_{\mathrm{ZZ}}$   & $95.70 \pm 2.08$ & $88.0 \pm 0.00$  & $91.68 \pm 0.95$ & $95.54 \pm 0.04$ & $85.6 \pm 0.80$  & $90.29 \pm 0.47$ \\
        && $U_{\mathrm{NC}}$  & $91.49 \pm 0.00$ & $86.0 \pm 0.00$  & $88.66 \pm 0.00$ & $95.56 \pm 0.00$ & $86.0 \pm 0.00$  & $90.53 \pm 0.00$ \\
        && $U_{\mathrm{NC10}}$& $94.52 \pm 1.55$ & $88.4 \pm 3.20$  & $91.30 \pm 1.08$ & $\mathbf{95.61 \pm 0.05}$ & $\mathbf{87.2 \pm 0.98}$  & $\mathbf{91.21 \pm 0.56}$ \\
        && $U_{\mathrm{NCL}}$ & $100.00 \pm 0.00$ & $85.6 \pm 1.50$  & $92.23 \pm 0.87$ & $93.59 \pm 0.08$ & $58.4 \pm 0.80$  & $71.92 \pm 0.63$ \\
        \cmidrule{2-9}
        &\multirow{4}{*}{$g_c$}
        & $U_{\mathrm{ZZ}}$   & $94.68 \pm 0.96$ & $80.0 \pm 0.00$  & $86.72 \pm 0.40$ & $\mathbf{87.17 \pm 2.41}$ & $\mathbf{96.8 \pm 2.40}$  & $\mathbf{91.67 \pm 0.16}$ \\
        && $U_{\mathrm{NC}}$  & $91.67 \pm 0.00$ & $88.0 \pm 0.00$  & $89.80 \pm 0.00$ & $88.68 \pm 0.00$ & $94.0 \pm 0.00$  & $91.26 \pm 0.00$ \\
        && $U_{\mathrm{NC10}}$& $85.54 \pm 3.53$ & $95.2 \pm 2.71$  & $90.00 \pm 0.84$ & $87.79 \pm 1.55$ & $86.0 \pm 1.79$  & $86.87 \pm 1.12$ \\
        && $U_{\mathrm{NCL}}$ & $92.17 \pm 1.98$ & $74.0 \pm 3.35$  & $82.00 \pm 1.34$ & $84.62 \pm 0.00$ & $88.0 \pm 0.00$  & $86.27 \pm 0.00$ \\
        \bottomrule
    \end{tabular}
    \caption{
Precision, recall, and F1-score for the experiments reported in Table~\ref{tab: combined_results}. For each dataset and interface model, the model with the highest accuracy is highlighted in bold. In cases where multiple models achieve the same accuracy, the one with the lower standard deviation is considered better and highlighted in bold.
    }
    \label{tab:combined_results_prf}
\end{table*}

\begin{table*}[ht]
\centering

\setlength{\tabcolsep}{8pt} 
\renewcommand{\arraystretch}{1.2}
\begin{tabular}{c|c|c|c|c|c}
    \toprule
    \textbf{Data} & \textbf{Interface Model} & \textbf{Classical Baseline Model} &
    \textbf{Precision (\%)} & \textbf{Recall (\%)} & \textbf{F1-score (\%)} \\
    \midrule
    \multirow{9}{*}{\makecell{CIFAR-10\\(Airplane–Automobile)}}
      & \multirow{3}{*}{$g_a$} & (1) & $\mathbf{70.0 \pm 18.4}$ & $\mathbf{93.8 \pm 5.7}$  & $\mathbf{78.5 \pm 10.9}$ \\
      &                        & (2) & $47.1 \pm 30.5$ & $78.8 \pm 44.1$ & $57.9 \pm 33.8$ \\
      &                        & (3) & $52.6 \pm 33.5$ & $73.8 \pm 42.2$ & $59.8 \pm 34.4$ \\
    \cmidrule{2-6}
      & \multirow{3}{*}{$g_b$} & (1) & $\mathbf{76.0 \pm 15.2}$ & $\mathbf{90.4 \pm 7.2}$  & $\mathbf{81.3 \pm 8.4}$ \\
      &                        & (2) & $10.0 \pm 22.4$ & $20.0 \pm 44.7$ & $13.3 \pm 29.8$ \\
      &                        & (3) & $55.2 \pm 11.6$ & $98.2 \pm 4.0$  & $69.9 \pm 7.2$ \\
    \cmidrule{2-6}
      & \multirow{3}{*}{$g_c$} & (1) & $\mathbf{79.5 \pm 2.5}$  & $\mathbf{84.6 \pm 4.7}$  & $\mathbf{81.9 \pm 2.3}$ \\
      &                        & (2) & $40.0 \pm 22.4$ & $80.0 \pm 44.7$ & $53.3 \pm 29.8$ \\
      &                        & (3) & $34.9 \pm 33.4$ & $58.0 \pm 53.1$ & $43.0 \pm 39.7$ \\
    \midrule
    \multirow{9}{*}{\makecell{CIFAR-10\\(Frog–Ship)}}
      & \multirow{3}{*}{$g_a$} & (1) & $\mathbf{92.4 \pm 3.0}$ & $\mathbf{93.0 \pm 3.1}$ & $\mathbf{92.6 \pm 1.7}$ \\
      &                        & (2) & $19.2 \pm 42.8$ & $18.2 \pm 40.7$ & $18.7 \pm 41.7$ \\
      &                        & (3) & $65.2 \pm 40.8$ & $75.8 \pm 42.6$ & $68.7 \pm 40.0$ \\
    \cmidrule{2-6}
      & \multirow{3}{*}{$g_b$} & (1) & $\mathbf{91.4 \pm 3.3}$ & $\mathbf{90.2 \pm 6.0}$ & $\mathbf{90.7 \pm 3.4}$ \\
      &                        & (2) & $40.0 \pm 22.4$ & $80.0 \pm 44.7$ & $53.3 \pm 29.8$ \\
      &                        & (3) & $39.2 \pm 40.4$ & $59.6 \pm 54.4$ & $46.1 \pm 43.8$ \\
    \cmidrule{2-6}
      & \multirow{3}{*}{$g_c$} & (1) & $\mathbf{93.2 \pm 2.2}$ & $\mathbf{90.6 \pm 5.7}$ & $\mathbf{91.7 \pm 2.0}$ \\
      &                        & (2) & $10.0 \pm 22.4$ & $20.0 \pm 44.7$ & $13.3 \pm 29.8$ \\
      &                        & (3) & $20.0 \pm 27.4$ & $40.0 \pm 54.8$ & $26.7 \pm 36.5$ \\
    \midrule
    \multirow{9}{*}{\makecell{Tiny ImageNet\\(School Bus–Maypole)}}
      & \multirow{3}{*}{$g_a$} & (1) & $\mathbf{88.2 \pm 15.4}$ & $\mathbf{88.0 \pm 3.5}$  & $\mathbf{87.4 \pm 8.7}$ \\
      &                        & (2) & $50.0 \pm 0.0$  & $100.0 \pm 0.0$ & $66.7 \pm 0.0$ \\
      &                        & (3) & $68.4 \pm 43.5$ & $76.8 \pm 43.0$ & $70.9 \pm 41.6$ \\
    \cmidrule{2-6}
      & \multirow{3}{*}{$g_b$} & (1) & $\mathbf{93.3 \pm 2.0}$  & $\mathbf{93.2 \pm 1.8}$  & $\mathbf{93.2 \pm 0.8}$ \\
      &                        & (2) & $10.0 \pm 22.4$ & $20.0 \pm 44.7$ & $13.3 \pm 29.8$ \\
      &                        & (3) & $20.0 \pm 27.4$ & $40.0 \pm 54.8$ & $26.7 \pm 36.5$ \\
    \cmidrule{2-6}
      & \multirow{3}{*}{$g_c$} & (1) & $\mathbf{83.9 \pm 6.2}$  & $\mathbf{85.2 \pm 8.8}$  & $\mathbf{84.1 \pm 3.1}$ \\
      &                        & (2) & $30.0 \pm 27.4$ & $60.0 \pm 54.8$ & $40.0 \pm 36.5$ \\
      &                        & (3) & $30.0 \pm 27.4$ & $60.0 \pm 54.8$ & $40.0 \pm 36.5$ \\
    \bottomrule
\end{tabular}

\caption{
Precision, recall, and F1-score for the experiments reported in Table~\ref{tab: classical_model_comparison}. For each dataset and the corresponding interface model, the classical baseline model with the highest accuracy is highlighted in bold.
}
\label{tab: classical_model_comparison_prf}
\end{table*}

\section{Experimental Setup for Autoencoder-Based Dimensionality Reduction} \label{Append: Performance Comparison with Autoencoder-Based Dimensionality Reduction}

The following details the experimental settings of the autoencoder-based dimensionality reduction experiments presented in Section~\ref{subsec: Performance Comparison with Autoencoder-Based Dimensionality Reduction}. All experiments were conducted on a 4-qubit system.

Autoencoders were trained to generate latent representations aligned with the input dimensionality required by the embedding circuits of models $g_a$ and $g_c$, which correspond to 8 and 24 dimensions, respectively. To ensure fairness, the autoencoder for $g_c$ was trained such that each channel was independently reduced to the designated dimensionality.

We employed a convolutional autoencoder to obtain latent embeddings. The encoder structure was designed to be identical to the classical component of CNQE for each model ($g_a, g_c$), and the decoder mirrored this architecture. The model was trained using mean squared error (MSE) loss and optimized with Adam at a learning rate of 0.01 for 2000 iterations, which was sufficient for the loss to converge. A batch size of 25 was used, with each mini-batch obtained by random resampling from the training set. Random parameter initialization was adopted. After training, only the encoder was retained to generate low-dimensional embeddings, while the decoder was used solely during the reconstruction phase.

Subsequently, QCNNs were trained under the same experimental settings as in our CNQE–QCNN model, and the entire process was repeated for five independent training runs.

\section{Experimental Setup for Noisy Simulation} \label{Append: Experimental Setup for Noisy Simulation}

The following details the experimental settings of the noisy simulations presented in Section~\ref{subsec: Performance under Noisy Simulation}.
 
We employed the Qiskit "FakeVigo" noise model, which emulates the noise characteristics of a superconducting qubit system.
The noise parameters (T1, T2, gate error rates, and measurement error rates) are listed in Table~\ref{tab:fakevigo_noise}.
The models evaluated in the noisy simulations were $g_a$ with ansatz $(b)$ and $g_c$ with $U_{\mathrm{NC}}$, and the fidelity loss was used for training.

\begin{table}[ht]
\centering
\renewcommand{\arraystretch}{1.2}
\begin{tabular}{l c}
\toprule
\textbf{Noise Parameter} & \textbf{Average Value} \\
\midrule
$T_1$ (relaxation time) & $108 \,\mu s$ \\
$T_2$ (dephasing time) & $70 \,\mu s$ \\
single-qubit gate error rate & $5.1 \times 10^{-4}$ \\
two-qubit gate error rate   & $8.8 \times 10^{-3}$ \\
measurement error rate & $3.34 \times 10^{-2}$ \\
\bottomrule
\end{tabular}
\caption{Average noise parameters of the Qiskit FakeVigo backend.}
\label{tab:fakevigo_noise}
\end{table}

Simulated noise was incorporated into both the CNQE and QCNN stages of the pipeline.
We evaluated its effect on the learning process using trace distance and classification accuracy.
Each experiment was repeated five times. To ensure fair comparison, all training hyperparameters were kept identical to those used in the noiseless experiments.



\clearpage

\begin{thebibliography}{10}

\bibitem{10.1038/s43588-022-00311-3}
M.~Cerezo, Guillaume Verdon, Hsin-Yuan Huang, Lukasz Cincio, and Patrick~J. Coles.
\newblock {Challenges and opportunities in quantum machine learning}.
\newblock {\em Nature Computational Science}, pages 1--10, 2022.

\bibitem{PhysRevLett.113.130503_QSVM}
Patrick Rebentrost, Masoud Mohseni, and Seth Lloyd.
\newblock Quantum support vector machine for big data classification.
\newblock {\em Phys. Rev. Lett.}, 113:130503, Sep 2014.

\bibitem{PhysRevLett.127.060503}
Ewin Tang.
\newblock Quantum principal component analysis only achieves an exponential speedup because of its state preparation assumptions.
\newblock {\em Phys. Rev. Lett.}, 127:060503, 2021.

\bibitem{doi:10.1126/science.abn7293}
Hsin-Yuan Huang, Michael Broughton, Jordan Cotler, Sitan Chen, Jerry Li, Masoud Mohseni, Hartmut Neven, Ryan Babbush, Richard Kueng, John Preskill, and Jarrod~R. McClean.
\newblock Quantum advantage in learning from experiments.
\newblock {\em Science}, 376(6598):1182--1186, 2022.

\bibitem{cerezo2020variational}
M.~Cerezo, Andrew Arrasmith, Ryan Babbush, Simon~C. Benjamin, Suguru Endo, Keisuke Fujii, Jarrod~R. {McClean}, Kosuke Mitarai, Xiao Yuan, Lukasz Cincio, and Patrick~J. Coles.
\newblock Variational quantum algorithms.
\newblock {\em Nature Reviews Physics}, 3(9):625--644, 2021.

\bibitem{cong_quantum_2019}
Iris Cong, Soonwon Choi, and Mikhail~D. Lukin.
\newblock Quantum convolutional neural networks.
\newblock {\em Nature Physics}, 15(12):1273--1278, December 2019.

\bibitem{farhi2014quantumapproximateoptimizationalgorithm}
Edward Farhi, Jeffrey Goldstone, and Sam Gutmann.
\newblock A quantum approximate optimization algorithm, 2014.

\bibitem{Peruzzo2014}
Alberto Peruzzo, Jarrod McClean, Peter Shadbolt, Man-Hong Yung, Xiao-Qi Zhou, Peter~J. Love, Al{\'a}n Aspuru-Guzik, and Jeremy~L. O'Brien.
\newblock A variational eigenvalue solver on a photonic quantum processor.
\newblock {\em Nature Communications}, 5(1):4213, Jul 2014.

\bibitem{Bravo_Prieto_2023}
Carlos Bravo-Prieto, Ryan LaRose, M.~Cerezo, Yigit Subasi, Lukasz Cincio, and Patrick~J. Coles.
\newblock Variational quantum linear solver.
\newblock {\em Quantum}, 7:1188, November 2023.

\bibitem{verdon2019quantumhamiltonianbasedmodelsvariational}
Guillaume Verdon, Jacob Marks, Sasha Nanda, Stefan Leichenauer, and Jack Hidary.
\newblock Quantum hamiltonian-based models and the variational quantum thermalizer algorithm, 2019.

\bibitem{kim2025expressivity}
Yujin Kim and Daniel~K Park.
\newblock Expressivity of deterministic quantum computation with one qubit.
\newblock {\em Physical Review A}, 111(2):022429, 2025.

\bibitem{Nguyen_2024}
Thien Nguyen, Tuomo Sipola, and Jari Hautamäki.
\newblock Machine learning applications of quantum computing: A review.
\newblock {\em European Conference on Cyber Warfare and Security}, 23(1):322–330, June 2024.

\bibitem{PERALGARCIA2024100619}
David Peral-García, Juan Cruz-Benito, and Francisco~José García-Peñalvo.
\newblock Systematic literature review: Quantum machine learning and its applications.
\newblock {\em Computer Science Review}, 51:100619, 2024.

\bibitem{e25020287}
Amine Zeguendry, Zahi Jarir, and Mohamed Quafafou.
\newblock Quantum machine learning: A review and case studies.
\newblock {\em Entropy}, 25(2), 2023.

\bibitem{senokosov2024quantum}
Arsenii Senokosov, Alexandr Sedykh, Asel Sagingalieva, Basil Kyriacou, and Alexey Melnikov.
\newblock Quantum machine learning for image classification.
\newblock {\em Machine Learning: Science and Technology}, 5(1):015040, 2024.

\bibitem{Preskill2018quantumcomputingin}
John Preskill.
\newblock Quantum {C}omputing in the {NISQ} era and beyond.
\newblock {\em {Quantum}}, 2:79, August 2018.

\bibitem{suzuki_analysis_2020}
Yudai Suzuki, Hiroshi Yano, Qi~Gao, Shumpei Uno, Tomoki Tanaka, Manato Akiyama, and Naoki Yamamoto.
\newblock Analysis and synthesis of feature map for kernel-based quantum classifier.
\newblock {\em Quantum Machine Intelligence}, 2(1):9, July 2020.

\bibitem{VQASVM}
Siheon Park, Daniel~K. Park, and June-Koo~Kevin Rhee.
\newblock Variational quantum approximate support vector machine with inference transfer.
\newblock {\em Scientific Reports}, 13(1):3288, 2023.

\bibitem{blank2020quantum}
Carsten Blank, Daniel~K Park, June-Koo~Kevin Rhee, and Francesco Petruccione.
\newblock Quantum classifier with tailored quantum kernel.
\newblock {\em npj Quantum Information}, 6(1):1--7, 2020.

\bibitem{park2021robust}
Daniel~K. Park, Carsten Blank, and Francesco Petruccione.
\newblock Robust quantum classifier with minimal overhead.
\newblock In {\em 2021 International Joint Conference on Neural Networks (IJCNN)}, pages 1--7, 2021.

\bibitem{PARK2020126422}
Daniel~K. Park, Carsten Blank, and Francesco Petruccione.
\newblock The theory of the quantum kernel-based binary classifier.
\newblock {\em Physics Letters A}, 384(21):126422, 2020.

\bibitem{grant_hierarchical_2018}
Edward Grant, Marcello Benedetti, Shuxiang Cao, Andrew Hallam, Joshua Lockhart, Vid Stojevic, Andrew~G. Green, and Simone Severini.
\newblock Hierarchical quantum classifiers.
\newblock {\em npj Quantum Information}, 4(1):65, December 2018.

\bibitem{PhysRevA.101.032308}
Maria Schuld, Alex Bocharov, Krysta~M. Svore, and Nathan Wiebe.
\newblock Circuit-centric quantum classifiers.
\newblock {\em Phys. Rev. A}, 101:032308, Mar 2020.

\bibitem{Helstrom1969}
Carl~W. Helstrom.
\newblock Quantum detection and estimation theory.
\newblock {\em Journal of Statistical Physics}, 1(2):231--252, Jun 1969.

\bibitem{Bae_2015}
Joonwoo Bae and Leong-Chuan Kwek.
\newblock Quantum state discrimination and its applications.
\newblock {\em Journal of Physics A: Mathematical and Theoretical}, 48(8):083001, jan 2015.

\bibitem{Nielsen:2011:QCQ:1972505}
Michael~A. Nielsen and Isaac~L. Chuang.
\newblock {\em Quantum Computation and Quantum Information: 10th Anniversary Edition}.
\newblock Cambridge University Press, New York, NY, USA, 10th edition, 2011.

\bibitem{hur2023neural}
Tak Hur, Israel~F. Araujo, and Daniel~K. Park.
\newblock Neural quantum embedding: Pushing the limits of quantum supervised learning, 2023.

\bibitem{Yamashita2018}
Rikiya Yamashita, Mizuho Nishio, Richard Kinh~Gian Do, and Kaori Togashi.
\newblock Convolutional neural networks: an overview and application in radiology.
\newblock {\em Insights into Imaging}, 9(4):611--629, Aug 2018.

\bibitem{schuld2021effect}
Maria Schuld, Ryan Sweke, and Johannes~Jakob Meyer.
\newblock Effect of data encoding on the expressive power of variational quantum-machine-learning models.
\newblock {\em Physical Review A}, 103(3):032430, 2021.

\bibitem{doi:10.1073/pnas.94.5.1634}
David~G. Cory, Amr~F. Fahmy, and Timothy~F. Havel.
\newblock Ensemble quantum computing by nmr spectroscopy.
\newblock {\em Proceedings of the National Academy of Sciences}, 94(5):1634--1639, 1997.

\bibitem{doi:10.1098/rsta.2011.0352}
Ben Criger, Gina Passante, Daniel Park, and Raymond Laflamme.
\newblock Recent advances in nuclear magnetic resonance quantum information processing.
\newblock {\em Philosophical Transactions of the Royal Society A: Mathematical, Physical and Engineering Sciences}, 370(1976):4620--4635, 2012.

\bibitem{NMRSimulator}
Ben Criger, Daniel Park, and Jonathan Baugh.
\newblock {\em Few-Qubit Magnetic Resonance Quantum Information Processors: Simulating Chemistry and Physics}, pages 193--228.
\newblock John Wiley \& Sons, Ltd, 2014.

\bibitem{Kyungdeock2015}
{Park, Kyungdeock}.
\newblock {\em Coherent control of nuclear and electron spins for quantum information processing}.
\newblock PhD thesis, 2015.

\bibitem{Lu2016}
Dawei Lu, Aharon Brodutch, Jihyun Park, Hemant Katiyar, Tomas Jochym-O'Connor, and Raymond Laflamme.
\newblock {\em NMR Quantum Information Processing}, pages 193--226.
\newblock Springer New York, New York, NY, 2016.

\bibitem{liu2025neuralquantumembeddingdeterministic}
Hongfeng Liu, Tak Hur, Shitao Zhang, Liangyu Che, Xinyue Long, Xiangyu Wang, Keyi Huang, Yu~ang Fan, Yuxuan Zheng, Yufang Feng, Xinfang Nie, Daniel~K. Park, and Dawei Lu.
\newblock Neural quantum embedding via deterministic quantum computation with one qubit, 2025.

\bibitem{holmes_2022}
Zoë Holmes, Kunal Sharma, M.~Cerezo, and Patrick~J. Coles.
\newblock {Connecting Ansatz Expressibility to Gradient Magnitudes and Barren Plateaus}.
\newblock {\em PRX Quantum}, 3(1):010313, 2022.

\bibitem{Sim_expressibility}
Sukin Sim, Peter~D. Johnson, and Alán Aspuru-Guzik.
\newblock Expressibility and entangling capability of parameterized quantum circuits for hybrid quantum-classical algorithms.
\newblock {\em Advanced Quantum Technologies}, 2(12):1900070, 2019.

\bibitem{Lee_2024}
Dongkeun Lee, Kyunghyun Baek, Joonsuk Huh, and Daniel~K Park.
\newblock Variational quantum state discriminator for supervised machine learning.
\newblock {\em Quantum Science and Technology}, 9(1):015017, nov 2023.

\bibitem{Havlicek2019}
Vojtech Havl{\'i}cek, Antonio~D. C{\'o}rcoles, Kristan Temme, Aram~W. Harrow, Abhinav Kandala, Jerry~M. Chow, and Jay~M. Gambetta.
\newblock Supervised learning with quantum-enhanced feature spaces.
\newblock {\em Nature}, 567(7747):209--212, 2019.

\bibitem{PhysRevLett.117.080501}
Michael~J. Bremner, Ashley Montanaro, and Dan~J. Shepherd.
\newblock Average-case complexity versus approximate simulation of commuting quantum computations.
\newblock {\em Phys. Rev. Lett.}, 117:080501, Aug 2016.

\bibitem{kandala_hardware-efficient_2017}
Abhinav Kandala, Antonio Mezzacapo, Kristan Temme, Maika Takita, Markus Brink, Jerry~M. Chow, and Jay~M. Gambetta.
\newblock Hardware-efficient variational quantum eigensolver for small molecules and quantum magnets.
\newblock {\em Nature}, 549(7671):242--246, September 2017.

\bibitem{toffano:hal-03042373}
Zeno Toffano and Alberto Ottolenghi.
\newblock {Quantum Entanglement and the Lorentz Group}.
\newblock In {\em {11th Colloquium on Quantum Engineering, Foundations \& Applications / Ing{\'e}nierie Quantique, des aspects Fondamentaux aux Applications -- IQFA, of the CNRS GDR-3322 , Universit{\'e} Grenoble Alpes (UGA)}}, Grenoble, France, December 2020. { CNRS GDR-3322 , Universit{\'e} Grenoble Alpes (UGA)}.

\bibitem{krizhevsky2009learning}
Alex Krizhevsky.
\newblock Learning multiple layers of features from tiny images.
\newblock pages 32--33, 2009.

\bibitem{endo2018practical}
Suguru Endo, Simon~C Benjamin, and Ying Li.
\newblock Practical quantum error mitigation for near-future applications.
\newblock {\em Physical Review X}, 8(3):031027, 2018.

\bibitem{temme2017error}
Kristan Temme, Sergey Bravyi, and Jay~M Gambetta.
\newblock Error mitigation for short-depth quantum circuits.
\newblock {\em Physical review letters}, 119(18):180509, 2017.

\bibitem{9226505}
Changjun Kim, Kyungdeock~Daniel Park, and June-Koo Rhee.
\newblock Quantum error mitigation with artificial neural network.
\newblock {\em IEEE Access}, 8:188853--188860, 2020.

\bibitem{Kurita2023synergeticquantum}
Tomochika Kurita, Hammam Qassim, Masatoshi Ishii, Hirotaka Oshima, Shintaro Sato, and Joseph Emerson.
\newblock Synergetic quantum error mitigation by randomized compiling and zero-noise extrapolation for the variational quantum eigensolver.
\newblock {\em {Quantum}}, 7:1184, November 2023.

\bibitem{Maciejewski2020mitigationofreadout}
Filip~B. Maciejewski, Zolt{\'{a}}n Zimbor{\'{a}}s, and Michal Oszmaniec.
\newblock Mitigation of readout noise in near-term quantum devices by classical post-processing based on detector tomography.
\newblock {\em {Quantum}}, 4:257, April 2020.

\bibitem{9142431}
Hyeokjea Kwon and Joonwoo Bae.
\newblock A hybrid quantum-classical approach to mitigating measurement errors in quantum algorithms.
\newblock {\em IEEE Transactions on Computers}, 70(9):1401--1411, 2021.

\bibitem{kim2022quantum}
Jihye Kim, Byungdu Oh, Yonuk Chong, Euyheon Hwang, and Daniel~K Park.
\newblock Quantum readout error mitigation via deep learning.
\newblock {\em New Journal of Physics}, 24(7):073009, 2022.

\bibitem{PhysRevApplied.17.014024}
Benjamin Lienhard, Antti Veps\"al\"ainen, Luke~C.G. Govia, Cole~R. Hoffer, Jack~Y. Qiu, Diego Rist\`e, Matthew Ware, David Kim, Roni Winik, Alexander Melville, Bethany Niedzielski, Jonilyn Yoder, Guilhem~J. Ribeill, Thomas~A. Ohki, Hari~K. Krovi, Terry~P. Orlando, Simon Gustavsson, and William~D. Oliver.
\newblock Deep-neural-network discrimination of multiplexed superconducting-qubit states.
\newblock {\em Phys. Rev. Applied}, 17:014024, Jan 2022.

\bibitem{Lee_2023}
Changwon Lee and Daniel~K Park.
\newblock Scalable quantum measurement error mitigation via conditional independence and transfer learning.
\newblock {\em Machine Learning: Science and Technology}, 4(4):045051, dec 2023.

\bibitem{AdaptVQE}
Harper~R. Grimsley, Sophia~E. Economou, Edwin Barnes, and Nicholas~J. Mayhall.
\newblock An adaptive variational algorithm for exact molecular simulations on a quantum computer.
\newblock {\em Nature Communications}, 10(1):3007, 2019.

\bibitem{QMI_semiagnostic}
M.~Bilkis, M.~Cerezo, Guillaume Verdon, Patrick~J. Coles, and Lukasz Cincio.
\newblock A semi-agnostic ansatz with variable structure for variational quantum algorithms.
\newblock {\em Quantum Machine Intelligence}, 5(2):43, 2023.

\bibitem{AdaptVQE-2}
Harper~R. Grimsley, George~S. Barron, Edwin Barnes, Sophia~E. Economou, and Nicholas~J. Mayhall.
\newblock Adaptive, problem-tailored variational quantum eigensolver mitigates rough parameter landscapes and barren plateaus.
\newblock {\em npj Quantum Information}, 9(1):19, 2023.

\bibitem{Wada_2024}
Kaito Wada, Rudy Raymond, Yuki Sato, and Hiroshi~C Watanabe.
\newblock Sequential optimal selections of single-qubit gates in parameterized quantum circuits.
\newblock {\em Quantum Science and Technology}, 9(3):035030, may 2024.

\bibitem{agarap2019deeplearningusingrectified}
Abien~Fred Agarap.
\newblock Deep learning using rectified linear units (relu), 2019.

\bibitem{hur2022quantum}
Tak Hur, Leeseok Kim, and Daniel~K Park.
\newblock Quantum convolutional neural network for classical data classification.
\newblock {\em Quantum Machine Intelligence}, 4(1):3, 2022.

\bibitem{kingma2017adam}
Diederik~P Kingma and Jimmy Ba.
\newblock Adam: A method for stochastic optimization.
\newblock {\em arXiv preprint arXiv:1412.6980}, 2017.

\bibitem{Qiskit}
{Qiskit contributors}.
\newblock Qiskit: An open-source framework for quantum computing, 2023.

\bibitem{bergholm2020pennylane}
Ville Bergholm, Josh Izaac, Maria Schuld, Christian Gogolin, M.~Sohaib Alam, Shahnawaz Ahmed, Juan~Miguel Arrazola, Carsten Blank, Alain Delgado, Soran Jahangiri, Keri McKiernan, Johannes~Jakob Meyer, Zeyue Niu, Antal Száva, and Nathan Killoran.
\newblock Pennylane: Automatic differentiation of hybrid quantum-classical computations.
\newblock {\em arXiv preprint arXiv:1811.04968}, 2020.

\end{thebibliography}
\end{document}